\documentclass[11pt]{article}
\pdfoutput=1

\usepackage{jheppub}
\usepackage[T1]{fontenc} 
\usepackage{subcaption}
\usepackage{longtable}
\usepackage[vcentermath]{youngtab}
\usepackage{array,multirow}
 \usepackage{multirow}
\usepackage{amssymb}
\usepackage{amsfonts}
\usepackage{amsbsy}
\usepackage{amsmath}
\usepackage{amsthm}
\usepackage{graphicx}
\usepackage{epstopdf}
\usepackage{color}
\usepackage{hyperref}
\usepackage{wasysym}
\usepackage{lscape}

\makeatletter
\g@addto@macro\bfseries{\boldmath}
\def\NAT@sort{\z@}
\makeatother

\def \be {\begin{equation}}
\def \ee {\end{equation}}
\def \bsp {\begin{split}}
\def \esp {\end{split}}
\def \bea {\begin{eqnarray}}
\def \eea {\end{eqnarray}}

\def\Z{\mathbb{Z}}
\def\F{\mathbb{F}}

\def\R{\mathbb{R}}

\def\P{\mathbb{P}}

\def\n3a{t}

\def\gso{{\mathfrak{so}}}
\def\gsu{{\mathfrak{su}}}
\def\gsp{{\mathfrak{sp}}}
\def\gf{{\mathfrak{f}}}
\def\gg{{\mathfrak{g}}}

\DeclareMathOperator{\Hom}{Hom}

\newcommand{\ds}{\ensuremath{\Delta}}
\newcommand{\dd}{\ensuremath{\nabla}}

\definecolor{Red}{rgb}{1.00, 0.00, 0.00}

\newcommand{\eq}[1]{(\ref{#1})}

\title{
Mirror symmetry and elliptic Calabi-Yau manifolds
}
\author{Yu-Chien Huang,}
\author{Washington Taylor}

\affiliation{Center for Theoretical Physics\\
Department of Physics\\
Massachusetts Institute of Technology\\
77 Massachusetts Avenue\\
Cambridge, MA 02139, USA}

\emailAdd{{\tt yc\_huang} {\rm at} {\tt mit.edu}}
\emailAdd{{\tt wati} {\rm at} {\tt mit.edu}}

\preprint{MIT-CTP-5068}

\abstract{
We find that for many Calabi-Yau threefolds with elliptic or genus one
fibrations mirror symmetry factorizes between the fiber and the base
of the fibration.  In the simplest examples, the generic CY elliptic 
fibration  over any toric base surface $B$ that supports an elliptic
Calabi-Yau threefold has a mirror that is an elliptic
fibration over a dual toric base surface $\tilde{B}$ that is related
through toric geometry to the line bundle $-6K_B$.  The Kreuzer-Skarke
database includes all these examples and gives a wide range of other
more complicated constructions where mirror symmetry also factorizes.
Since recent evidence suggests that most Calabi-Yau threefolds are
elliptic or genus one fibered, this points to a new way of
understanding mirror symmetry that may apply to a large fraction of
smooth Calabi-Yau threefolds.  The factorization structure identified
here can also apply for Calabi-Yau manifolds of higher dimension.}

\begin{document}
\maketitle

\flushbottom


\section{Introduction}
\label{sec:intro}

Calabi-Yau manifolds have been a major subject of study in mathematics
and physics over the last three decades, following the realization
that these geometries can be used to compactify string theory in a way
that preserves supersymmetry \cite{chsw}.  One of the most intriguing
aspects of Calabi-Yau threefolds is the existence of an equivalence
known as mirror symmetry
(see e.g. \cite{mirror-book}) that relates the physics of a type IIA string
compactification on a Calabi-Yau threefold $X$ to that arising from
a type IIB string
compactification on a mirror threefold $\tilde{X}$.  One of the first
important clues to mirror symmetry was the observation\cite{cls, Batyrev:1994hm} that the
Calabi-Yau threefolds generated by hypersurfaces in toric varieties
have the property
that for every Calabi-Yau threefold $X$ with Hodge numbers $h^{1,1}
(X), h^{2,1} (X)$ there is a corresponding mirror threefold
$\tilde{X}$ with Hodge numbers $h^{1,1} (\tilde{X}) = h^{2,1} (X),
h^{2,1} (\tilde{X}) = h^{1,1} (X)$.  
The largest known set of Calabi-Yau threefolds are constructed from
 the class of over 400 million reflexive 4D polytopes found by
Kreuzer and Skarke \cite{Kreuzer:2000xy, database}, and exhibit this
mirror symmetry structure.

More recently, an increasing body
of evidence \cite{Hodge, Candelas-cs, Gray-hl, Johnson-WT,
  Anderson-ggl, aggl-2, aggl-3, agh-non-simply, Huang-Taylor-1,
  Huang-Taylor-2} suggests that a large fraction of known Calabi-Yau
threefolds have the property that they can be described as genus one
or elliptic fibrations over a complex two-dimensional base surface.
We recently showed that this is true of all but at most 4 Calabi-Yau
threefolds in the Kreuzer-Skarke database having one or the other
Hodge number $h^{2,1}, h^{1,1}$ at least 140, and that at small
$h^{1,1}$ the fraction of polytopes in the Kreuzer-Skarke database
that lack an obvious elliptic or genus one fibration decreases roughly
as $0.1 \times 2^{5 - h^{1,1}}$.  In this paper we show that the
structure of these fibrations gives a natural way of ``factorizing''
mirror symmetry for many elliptic and genus one fibered toric
hypersurface Calabi-Yau threefolds, so that the fiber of $X$
determines the fiber of the mirror threefold
$\tilde{X}$, and the base and fibration structure of $X$ determine the
base of $\tilde{X}$.  A key aspect of this factorization involves the
observation that a simple additional condition on the fibration
structure of an elliptic toric hypersurface Calabi-Yau threefold $X$
implies that the mirror $\tilde{X}$ is also elliptic and has a mirror elliptic
fiber characterized by a 2D polytope dual  to the one that
contains the genus one or elliptic fiber of $X$.  Such mirror fibers
were also studied in the related context of K3 surfaces and
heterotic/F-theory duality in \cite{Avram:1996pj, Berglund-Mayr,
  Grassi-P, Candelas-cs, Cvetic-gp}, and in
the context of elliptic fibers for F-theory in \cite{Klevers-16}.

The outline of this paper is as follows: In \S\ref{sec:fibrations}, we
review some basic aspects of toric hypersurface Calabi-Yau manifolds and
elliptic and genus one fibrations; we then describe in general the way
in which mirror symmetry can factorize for toric hypersurface
Calabi-Yau manifolds.  In \S\ref{sec:generic}, we consider the
simplest examples of this factorization: when $X$ is the generic
CY elliptic fibration over any toric base surface $B$ that supports an elliptic
Calabi-Yau threefold, the mirror $\tilde{X}$ has a simple description as an
elliptic fibration over a dual base
$\tilde{B}$ that has a simple description in
toric geometry in terms of $B$.
 In \S\ref{sec:further-examples}, we consider some further examples,
including Weierstrass/Tate tunings of generic models over a toric
base, ``stacked'' fibrations with non-generic fiber types, and
analogous factorization for elliptic Calabi-Yau fourfolds.  We
conclude in \S\ref{sec:conclusions} with a summary and some open
questions.

\section{Toric hypersurface Calabi-Yau manifolds, fibrations and
  mirror symmetry}
\label{sec:fibrations}

In this section we review some basic aspects of toric hypersurface
Calabi-Yau threefolds and fibrations, and we describe the basic
framework of mirror symmetry factorization that applies for many
elliptic and genus one fibered toric hypersurface Calabi-Yau manifolds.  Much
of the material
reviewed in the first part of this section is covered in more
detail in the papers \cite{Huang-Taylor-1, Huang-Taylor-2}.

\subsection{Toric hypersurfaces and fibrations}
\label{sec:intro-fibrations}

A broad class of Calabi-Yau manifolds can be described as
hypersurfaces in toric varieties following the approach of Batyrev
\cite{Batyrev}.
A {\it lattice polytope} 
$\dd$
is defined to be the set of lattice points in
$N =\Z^n$ that are contained within the convex hull of a finite set of
vertices $v_i \in N$.
 The dual of a polytope $\nabla$ is defined to be 
\begin{equation}
\nabla^*=\{u\in M_\R=  M\otimes \R: \langle u,v\rangle\geq-1, \forall v\in \nabla\},
\label{dual}
\end{equation}
where $M = N^*=\Hom(N,\Z)$ is the dual lattice.  A lattice polytope
$\nabla\subset N$ containing the origin is {\it reflexive} if its dual
polytope is also a lattice polytope.
For any reflexive polytope, the origin is the unique interior point.

When $\dd$ is reflexive, we denote the dual polytope by $\ds =\dd^*$.
The elements of the dual polytope $\ds$ can be associated with
monomials in a section of the anti-canonical bundle of a toric variety
associated to $\dd$.   A section of this bundle defines a hypersurface
in the toric variety associated to $\dd$; this hypersurface is a Calabi-Yau manifold of dimension $n-1$.
The polytopes $\dd$ and $\ds$ describe  toric hypersurface
Calabi-Yau manifolds that are related by mirror symmetry \cite{Batyrev:1994hm}.
As $\dd$ and $\ds$ are a
  pair of 4D reflexive polytopes, there is a one-to-one
  correspondence between $l$-dimensional faces $\theta$ of $\ds$ and
  $(4-l)$-dimensional faces $\tilde{\theta}$ of $\dd$ related by the
  dual operation 
\begin{equation}
\theta^*=\{y\in \dd, \langle y, pt\rangle=-1 \rvert \text{ for all $pt$ that are vertices of $\theta$}\} \,.\label{eq:dual-faces}
\end{equation}
For the CY associated with $\dd$, the Hodge numbers are given by
\begin{eqnarray}
\label{latth21}
&&h^{2,1}=\text{pts}(\ds)-\sum_{\theta\in F^\ds_3}	\text{int}(\theta)+\sum_{\theta\in F^\ds_2} \text{int}(\theta)\text{int}({\theta}^*)-5,\\
\label{latth11}
&&h^{1,1}=\text{pts}(\dd)-\sum_{\tilde{\theta}\in F_3^\dd}	\text{int}(\tilde{\theta})+\sum_{\tilde{\theta}\in F_2^\dd} \text{int}(\tilde{\theta})\text{int}(\tilde{\theta}^*)-5,
\end{eqnarray}
where $\theta$ are faces of $\ds$,  $\tilde{\theta}$ are faces of
$\dd$, $F^{\dd\slash\ds}_l$ denotes the set of  $l$-dimensional faces
of $\dd$ or $\ds$ ($l<n$), and $\text{pts}(\dd\slash\ds):=$ number of
  lattice points of $\dd$ or $\ds$,
  int$(\theta\slash\tilde{\theta}):=$ number of lattice points
  interior to $\theta$ or $\tilde{\theta}$. 
The correspondence (\ref{eq:dual-faces}) makes the duality between the Hodge number formulae
manifest.

When the polytope $\dd$ has a 2D subpolytope $\dd_2$ that is also
reflexive, the associated Calabi-Yau manifold has a genus one
fibration \cite{Kreuzer:1997zg}.\footnote{In \cite{Rohsiepe} it is argued
  that in some cases the condition for a fibration is more subtle; in
  particular, the clearest way of constructing a fibration uses a
  toric morphism that  must be compatible with a triangulation of
  $\nabla$, which may not be possible in some cases, particularly in
higher dimensions.  
We do not analyze the detailed structure of triangulations in this
paper; for the simple cases considered here there does not seem to be
any obstruction to the existence of a triangulation giving a toric
morphism compatible with the fibration, but this in principle should
be checked in detail, particularly for higher-dimensional varieties
where the triangulation of the base is not unique.
}
This fibration is characterized by a projection  $\pi$ on $N=\Z^n$ that
maps the fiber $\dd_2$ to 0.  For a 4D polytope $\dd$, the base  $B$ of the
fibration is described by the 2D toric variety associated with the set
of primitive rays in the image of
$\nabla$ under the projection $\pi:\Z^4
\rightarrow\Z^2$, where a primitive ray is one that is not an integer
multiple of another element of the lattice.  Since the base in this
case is two-dimensional, the triangulation in the resulting toric
variety is uniquely determined by the ordering of the rays; in higher
dimensions, there may be many distinct triangulations possible.

There are 16 distinct reflexive 2D polytopes, listed in
Appendix~\ref{sec:appendix-fibers}. The structure of the genus one and elliptic fibrations
associated with each of these 16 fibers is studied in some detail in
\cite{Bouchard-Skarke} and in
the F-theory context in \cite{Braun:2011ux, BGK-geometric,
  Klevers-16, Huang-Taylor-2}.
As discussed in these papers, all fibers other than $F_1, F_2, F_4$
contain at least one curve of self-intersection $-1$; such a
curve gives rise
to a global
section so that the fibration is elliptic and not just genus one.
Associated with each of the 16 reflexive 2D polytopes $F =\dd_2 =
F_i$ is a  dual fiber $ \tilde{F}$, given by $\tilde{F_i} =\ds_2 =
F_{17-i}$ for all $i$ except $i = 7, 8, 9, 10$ in which cases
$\tilde{F_i}  \cong F_i$ under a linear change of coordinates.  
We will
refer to $\tilde{F}$ as the {\it mirror fiber}
of $F$.  The anti-canonical hypersurfaces in $F, \tilde{F}$ represent
a mirror pair of 1D Calabi-Yau varieties (genus one curves).
Aspects of
these dual fibers and associated mirror curves have previously been
encountered and studied in the contexts of K3 fibrations,
heterotic/F-theory duality and F-theory fibers
in \cite{Avram:1996pj, Berglund-Mayr, Grassi-P, Candelas-cs, Klevers-16, Cvetic-gp}.

When a polytope $\nabla$ admits a toric fibration of this kind,
the lattice points in the dual polytope $\Delta$ can be associated
with monomials in various line bundles over the base $B$.
We can choose a coordinate system on $N$ so that the vertices of the
fiber $ F =\dd_2$ lie in the plane $(0, 0; \cdot, \cdot)$.
Each lattice point $v \in \nabla$ can then be represented in the
form
\begin{equation}
v =(v_1, v_2; v_3, v_4)= (v^{(I)}; v^{(II)})\,,
\label{eq:ray}
\end{equation}
where the first two coordinates 
$v^{(I)} = (v_1, v_2)$
correspond to the base direction, and the last two coordinates
  $v^{(II)}\equiv(v_3,v_4)$
 correspond to
the toric fiber direction.
For each point $v^{F} \in\dd_2$, the point $v= (0; v^F)$ lies in
$\dd$.\footnote{Note that $v^{(II)}$ in the fiber direction may lie outside the
  fiber $\nabla_2$ for a general point
$(v^{(I)}; v^{(II)})$ in $\nabla$. In fact, as we discuss further
  later in the paper, this has to be
  the case for some lattice points in $\nabla$ when the dual polytope
  $\Delta$ is not a fibered polytope.}
The primitive rays
defining the base as a toric variety are those that are not integer
multiples of another ray,
\begin{equation}
\{v^{B}\}=\{v^{(I)} =(v_1,v_2)/({\rm GCD} (v_1, v_2)),
\exists v^{(II)}: v= (v^{(I)}, v^{(II)}) \in\dd\} \,.
\label{eq:base}
\end{equation} 
The existence of the projection $\pi:v= (v_1, v_2; v_3,
v_4)\rightarrow (v_1, v_2; 0, 0)$
taking $\nabla_2 \rightarrow 0$ is equivalent to the condition
 that there is a projection on the dual lattice $\rho:  m =
(m_1, m_2;
m_3, m_4) \rightarrow (0, 0; m_3, m_4)$ that maps $\ds$ to the mirror fiber
$\tilde{F}=\ds_2$.\footnote{This can be easily proven as follows:
Given the projection $\pi$, we know that the fiber $\dd_2$
lies in the plane $(0, 0; \cdot, \cdot)$.  This implies that any $m =
(m_1, m_2;m_3,m_4)\equiv (m^{(I)}; m^{(II)})$ satisfies $m^{(II)} \cdot v^F
\geq -$1, 
$\forall
v^F \in\dd_2$, which implies $m^{(II)}\in\Delta_2$, i.e., the existence
of the projection $\rho$ onto $\Delta_2$.  To see the existence of the
fiber given the projection $\rho$, every point in the form $v=(0,0;
v^F)$, where $v^F \in\dd_2$,
 satisfies $v\cdot m \geq -1, \forall m\in \Delta$
since
 $v^F\cdot m^{(II)}\geq -1, \forall m^{(II)}\in \Delta_2$; 
we therefore have $\nabla_2 \subset \nabla$, and the projection taking
$\nabla_2 \rightarrow 0$ takes the
form $\pi$.}
For each $m^{(II)} \in \ds_2$, the set of lattice points in $\ds$
that map under $\rho$ to $m^{(II)}$ can be thought of as monomials that are
sections of a specific line bundle over $B$.
Each ray of the form (\ref{eq:ray}) satisfies the condition $v\cdot m
\geq -1$, which implies $(v_1,v_2) \cdot (m_1,m_2) \geq -1 -v^{(II)} \cdot m^{(II)}$.  When $v^B=(v_1,v_2)$ is a
primitive ray and corresponds to a toric curve $C$ in the base $B$, this
means that $m^{(I)}= (m_1,m_2)$ is a section of a line bundle that can vanish to order
$v^{(II)} \cdot m^{(II)} + 1$ on $C$.

The simplest examples of the utility of these conditions can be seen
in polytopes that have the ``stacked'' form described in
\cite{Skarke-intro, Huang-Taylor-1, Huang-Taylor-2}, where there is a fixed lattice
point $v_s \in\dd_2$ so that
for every ray $v^B$ in the base there exists a ray  of the form
$(v^B;v_s)\in\dd$ for that particular lattice point $v_s$.
In these cases, the monomials over $m^{(II)}$ represent sections of the line
bundle ${\cal O} (-nK_B)$, where $n = 1 + v^{(II)} \cdot m^{(II)}$
and $-K_B$ is the anti-canonical class of the base.
In particular, when the fiber is $F_{10} =\P^{2, 3, 1}$, and $v_s = (-3,
-2)$ the resulting ``standard stacking'' form gives monomials over the
points in $\ds_2$ that naturally describe the general (Tate) form of
the Weierstrass model for an elliptic fibration, and can be described as sections of
${\cal O} (-nK_B)$, with $n = 1, 2, 3, 4, 6$.

\subsection{Factorization of mirror symmetry}
\label{sec:2.2}
From the preceding characterization of polytope fibrations, it clearly
follows that under certain circumstances when the polytope $\dd$ has a
subpolytope $\dd_2= F_i$ that gives a genus one fibration of the associated
Calabi-Yau hypersurface $X$
the polytope $\ds$ will also have a
subpolytope $\ds_2 = \tilde{F_i}$ that gives a genus one fibration of
the mirror Calabi-Yau $\tilde{X}$.
This will occur whenever there is a coordinate system such that the point $(0, 0; m^{(II)})$ is a point  in $\ds$ for all $m^{(II)}\in \Delta_2$. A
necessary and sufficient condition for this to occur is that there
exist a coordinate system so that every lattice point in $\dd$ can
simultaneously
be
put in the form (\ref{eq:ray}), with $v^{(II)} =v^F \in\dd_2$ (the values of $v^F$
need not be the same for different lattice points in $\dd$ but they must all
lie in $\dd_2$), i.e. that there is
a projection in the space $N$ onto the fiber polytope $\dd_2$.\footnote{This condition was  encountered in the context of K3 surfaces
and heterotic F-theory duality in
several earlier papers; mirror K3 fibrations were described in terms
of slices and projections in \cite{Avram:1996pj, Candelas-cs}, and described
 in
terms of  symplectic cuts
in \cite{Grassi-P}, motivated by some examples found by Candelas.
This type of construction has been further used in studying mirror
symmetry of $G_2$ manifolds \cite{Braun:2017ryx}.
The strong prevalence of mirror symmetric pairs of K3-fibered CY3s observed in \cite{Candelas-cs} is closely related to the prevalence of mirror symmetric pairs of elliptically-fibered CY3s studied here; indeed, many of the examples we consider here are also K3 fibered.
}

In any situation where these conditions hold, we have a mirror pair of
Calabi-Yau manifolds
\footnote{\label{ft1}The polytopes in the KS database \cite{database} are
  associated with the monomial polytopes $\ds$. Given a mirror pair of
  fibered polytopes $\nabla, \Delta$, where $\nabla$ is the ``fan polytope''
  associated with the fibration $X$ with the Hodge numbers ($h^{1,1}(X)$,
  $h^{2,1}(X)$), it is the $\Delta$ polytope associated with the data
  M:$\#$ lattice points, $\#$ vertices (of $\ds$) N:$\#$ lattice
  points, $\#$ vertices (of $\dd$) H: $h^{1,1}(X)$, $h^{2,1}(X)$ that is
listed on
  the website. Then $\Delta$ is the fan  polytope associated
  with the fibration $\tilde{X}$ with the Hodge numbers ($h^{1,1}(\tilde X)=h^{2,1}(X)$,
  $h^{2,1}(\tilde X)=h^{1,1}(X)$)
  While the lattice points in the fan polytope $\nabla$
  associated with the fibration $X$ are denoted by $v$, when $\Delta$
  is viewed as the fan polytope associated with the fibration $\tilde
  X$, we sometimes denote the lattice points in $\Delta$ by $w$, while we use the
  same symbol $m$ to denote the lattice points in either $\Delta$ or
  $\nabla$ in the cases when they are used as the monomial polytope.}
 $X, \tilde{X}$, each of
which is elliptically or genus one fibered.  Furthermore in the toric
presentation, the 2D toric fibers associated with the elliptic or
genus one fibers of $X, \tilde{X}$ themselves have mirror hypersurface
curves, with $F = F_i, \tilde{F} = \tilde{F_i}$.  We refer to this
situation as a ``factorization'' of mirror symmetry for elliptic
Calabi-Yau manifolds.

This kind of factorization is really in some sense a
semi-factorization.  In particular, the relationship between the base
$B$ of the elliptic fibration of $X$ and the base $\tilde{B}$ of the
mirror fibration depends upon the ``twist'' of the fibration of $X$
encoded in the specific way in which the rays of the base lie over the
fiber  in (\ref{eq:ray}).  In general, this relationship can be
rather complex, though it can always be determined from the condition
described in \S\ref{sec:intro-fibrations} that implies that
each primitive base ray in the mirror is associated with a section
$m^{(I)}$ of a line bundle whose degree of vanishing on the associated
curve is constrained by the set of inner products $v^{(II)} \cdot m^{(II)}$.

In many  cases, the structure is particularly simple
and the base $\tilde{B}$ of the mirror elliptic fibration can be
associated with a line bundle ${\cal O} (-nK_B)$ for a fixed $n$, so
that $\tilde{B}$ can be
read
off in a simple way from $B$.  In particular, this occurs when all the
rays of the base  are stacked over a particular point $v_s \in\dd_2$ in
the fiber, and there are no rays in $\dd$
(associated with ``tops'' \cite{tops})  representing rays in the base over other
fiber points
that
impose extra constraints on the points in the mirror polytope.  
In this case, the monomials in the dual polytope
$\tilde{B}$ can be
associated with sections of line bundles ${\cal O} (- n K_B)$, with $n
= 1+ v_s \cdot m^{(II)}$, and the base
$\tilde{B}$ can be associated with a polytope built
from the primitive rays in the set of points in the 2D polytope
associated with $- n K_B$ with the largest value of $n$ realized from
the points $m^{(II)} \in\ds_2$:
\begin{eqnarray}
\{w_i^{\tilde B}\}=\{ w =(w_1,w_2)\rvert &&\text{GCD}(w_1, w_2)=1, 
w \cdot v^{B} \geq -n_{v_s} \;\forall v^B \in \Sigma_B
\},
\label{eq:dual-base}
\end{eqnarray}
where $\Sigma_B$ is the toric fan for
$B$ and $n_{v_s}={\rm max}(\{(1+v_s\cdot m^{(II)})\rvert
m^{(II)}\in\Delta_2\})$. In Appendix~\ref{sec:appendix-fibers} we illustrate for each $v_s \in\dd_2$
the maximum value of $1 + v_s \cdot m^{(II)}$.
Note that unlike in \eq{eq:base}, where we are dealing with
projections of 4D rays, when an integer multiple $kw$ of a primitive 2D
vector $w$ satisfies $kw \cdot v^B \geq -n$ then the vector $w$ also
satisfies this condition, so we can construct all primitive vectors by
simply taking those with unit GCD on the coordinates.
The simplest cases in which (\ref{eq:dual-base}) applies is for the
``standard stacking'' $F_{10}$ fiber constructions associated with
the generic Tate form for an elliptic fibration over a toric base, as
discussed at the end of \S\ref{sec:intro-fibrations}, in which case
$n_{v_s}= 6$.  We describe a number of examples of this type in
\S\ref{sec:generic}.  When there is a further tuning of the monomials
in $\Delta$, associated with a  nontrivial ``top'' in $\dd$, the
construction of the mirror base $\tilde{B}$ is similar but  depends on
the tuning, as we discuss in more detail in \S\ref{sec:further-examples}

All of the analysis just outlined is equally relevant taking mirror
symmetry the other way.  Starting with $\Delta$,
the
base $B$ can similarly in the complementary cases, such as  when
$\Delta$ is a standard stacking associated with a generic elliptic
fibration over $\tilde{B}$, 
be calculated using (\ref{eq:dual-base}) from
$\mathcal{O}(-n_{w_s}K_{\tilde B})$, where $w_s\in \Delta_2$ is the
stacking point of the stacked polytope $\Delta$, and $n_{w_s}={\rm
  max}(\{(1+w_s\cdot m^{(II)})\rvert m^{(II)}\in\nabla_2\})$, and the
same kind of generalization is used when there is a tuning of the monomials
in $\nabla$. In the
remainder of the paper we consider 
primarily examples of this type, where there is a clear factorization
of the mirror symmetry that allows a ready identification of both the
fiber and the base of the mirror fibered Calabi-Yau variety.  We leave
a further analysis of more general cases for future work.

\section{Generic Calabi-Yau elliptic fibrations over toric base surfaces}
\label{sec:generic}

In this section we consider the simplest and perhaps clearest class of
examples of the factorization of mirror symmetry described above:
Calabi-Yau threefolds $X$ that are generic CY elliptic fibrations over a
toric base surface $B$.
The closely related class of threefolds resulting from tuned Tate-form
Weierstrass models over a toric base provide a larger class of
examples where the mirror symmetry factorization can be understood
easily; examples of this broader class  are given in 
\S\ref{sec:Hirzebruch},
\S\ref{sec:tuning}, and \S\ref{sec:tuning-2}.

\subsection{General case}

It was shown in \cite{toric} that there are 61,539 toric bases that
support an elliptic Calabi-Yau threefold with a smooth
resolution.\footnote{Note that these bases include those with curves
  of self-intersection $-9,-10, -11$.  Such bases can be blown up at
  non-toric points to achieve a smooth 
flat fibration Calabi-Yau resolution; Calabi-Yau threefolds can also
be realized as non-flat fibrations over these bases.
This
  technicality is handled automatically through the resolution process for the toric hypersurface
  Calabi-Yau threefolds in the Kreuzer-Skarke database
  \cite{Huang-Taylor-1} (see also examples in Table \ref{f:Hirzebruch-bases-nontoric} and an example in \S\ref{sec:tuning-3}.)}
The Hodge numbers of the generic elliptic fibrations over all these
bases were analyzed in \cite{Hodge}.  It was shown in
\cite{Huang-Taylor-1} that for each toric base $B$, a reflexive 4D
polytope in the Kreuzer-Skarke database can be constructed by starting
with a ``standard stacking'' polytope defined by the convex hull of
the set of points of the form (\ref{eq:ray}), where the first two coordinates are taken
across all toric rays $v^B$ in the fan of $B$, and the last two
coordinates
 $v_s = (-3, -2)$ correspond to a
vertex of the fiber $F = F_{10} = \P^{2,3,1}$, and then taking the
``dual of the dual'' of the resulting polytope.  In the simplest
cases, where $B$ only has curves of self-intersection $-n$ where $n |
12$, the initial polytope is already reflexive.
In all these cases, there is a corresponding reflexive 4D polytope in
the Kreuzer-Skarke database that has an explicit $F_{10}$ fiber.

\begin{figure}
\begin{center}
\includegraphics[width=8cm]{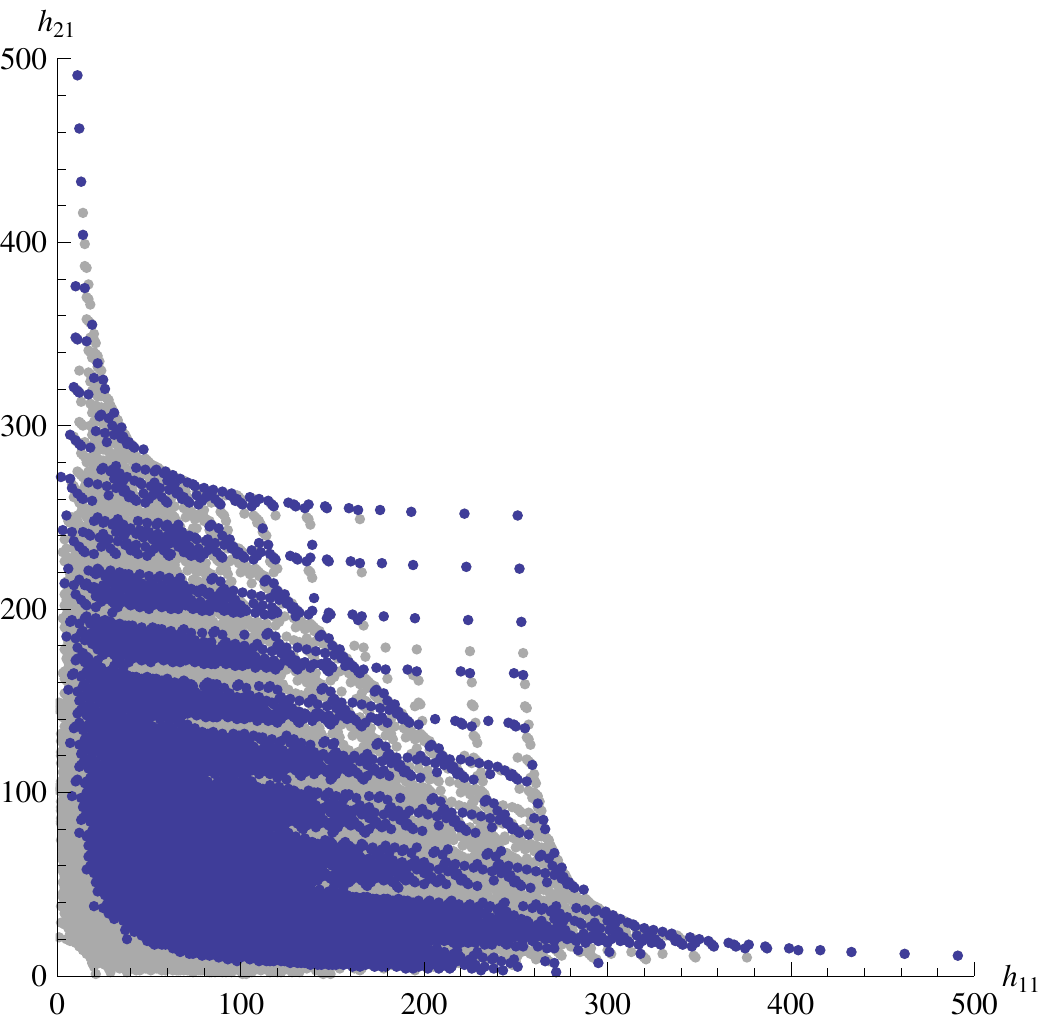}
\end{center}
\caption[x]{\footnotesize The Hodge numbers (blue points) of
  Calabi-Yau threefolds $X$ formed as generic elliptic fibrations over
  toric base surfaces.  All these examples have a particularly simple
  form of factorized mirror symmetry in which $X$ and $\tilde{X}$ both
  have elliptic fibers described as anti-canonical curves in the toric
  fiber $F_{10} =\P^{2, 3, 1}$ and the mirror base $\tilde{B}$ is
  constructed as a toric variety from the monomials in the line bundle
  ${\cal O} (-6K_B)$, where $B$ is the base of the elliptic fibration
  of $X$.  Gray points are Hodge numbers in the full KS database.  }
\label{f:Hodge-generic}
\end{figure}

For each of these generic elliptic fibrations over a toric base, the
dual polytope $\Delta$
associated with the mirror Calabi-Yau threefold  $\tilde{X}$
contains lattice points that can be
interpreted as sections of line bundles ${\cal O} (- n K_B)$ with $n =
1, 2, 3, 4, 6$.  These correspond to the coefficient polynomials in
the ``Tate form'' of a Weierstrass model\footnote{The term  ``Tate
  form'' is used often by physicists for this general form of the
  Weierstrass model because of its use in the context of F-theory in
  the Tate algorithm for constructing models with a particular desired
  gauge group.}
\begin{equation}
 y^2 + a_1 y x + a_3 y = x^3 + a_2 x^2 + a_4 x + a_6 \,,
 \label{eq:Tate}
\end{equation}
where $a_n$ is a section of ${\cal O} (- n K_B)$.  Since the origin is
contained in each of the 2D sets of points over each $m^{(II)} \in
\Delta_2$, for all these models the dual polytope $\Delta$ has a
subpolytope $\tilde{F}=\Delta_2$ lying on the plane $m_1=m_2=0$, 
so the
resulting mirror $\tilde{X}$ also has an elliptic fiber given by an
anti-canonical curve in the toric fiber $\tilde{F}\cong F = F_{10}$.
These correspond to one of the simplest classes of elliptic toric
hypersurface Calabi-Yau threefolds with a simple and manifest
factorization of mirror symmetry.  In these cases, the dual polytope
has a base that is described as a toric variety by the set of
primitive rays associated with the monomials in ${\cal O} (- 6 K_B)$,
which lie over the point $m^{(II)} = (-1, -1) \in \Delta_2$.
The Hodge numbers of the generic elliptic fibrations over toric bases
are plotted in Figure~\ref{f:Hodge-generic} \cite{Hodge}; these
include many of the largest Hodge number pairs in the KS
database.\footnote{The simplest subset
of these cases, where both sides of the mirror symmetry
are generic elliptic fibrations without
tuning, and some of the patterns appearing in these cases, were noted in the context of an earlier project with Braun and
Wang \cite{btw-unpublished}.}

In the remainder of this section we describe some examples of these
factorized mirror pairs explicitly.  From this construction, it is
clear that similarly, an arbitrary Tate tuning of the generic elliptic
fibration over a toric base $B$, which is realized by a reduction in
the set of monomials in $\Delta$ and an increase in the set of rays in
$\nabla$ (often described in the language of ``tops'' \cite{tops}), as
described in more detail in \cite{Huang-Taylor-1}, will also lead to a
mirror pair of Calabi-Yau threefolds that are both elliptically
fibered with the self-dual toric 2D fiber type
$F_{10}$. 
 In general,  tuning the fibration $X$ will reduce the
  size of the polytope associated with the mirror base $\tilde{B}$, 
  which will then be described by a toric fan that
 contains as rays only a subset of the primitive rays in
 $-6K_B$.\footnote{Note that  in some cases when all monomials over
  some points $m^{(II)} \in \Delta_2$ are set to vanish, this
  furthermore
will correspond
to reducing the size of $\Delta_2$, with a corresponding increase in
the size of $\nabla_2$; for example, setting $a_6 = 0$ in the Tate
model changes the dual fiber to $\tilde{F}_{13}$, so that the fiber of
$\nabla$ acquires two additional points and becomes $F_{13}$.}  We describe
 examples of such tunings in  \S\ref{sec:tuning}-\ref{sec:tuning-3}.

\subsection{Example: generic elliptic fibration over $\P^2$ (Hodge
  numbers (2, 272))}
\label{sec:example-p2}

\begin{figure}
\centering
\begin{picture}(200,240)(- 100,- 120)
\put(-120,0){\makebox(0,0){
    \includegraphics[width=2cm]{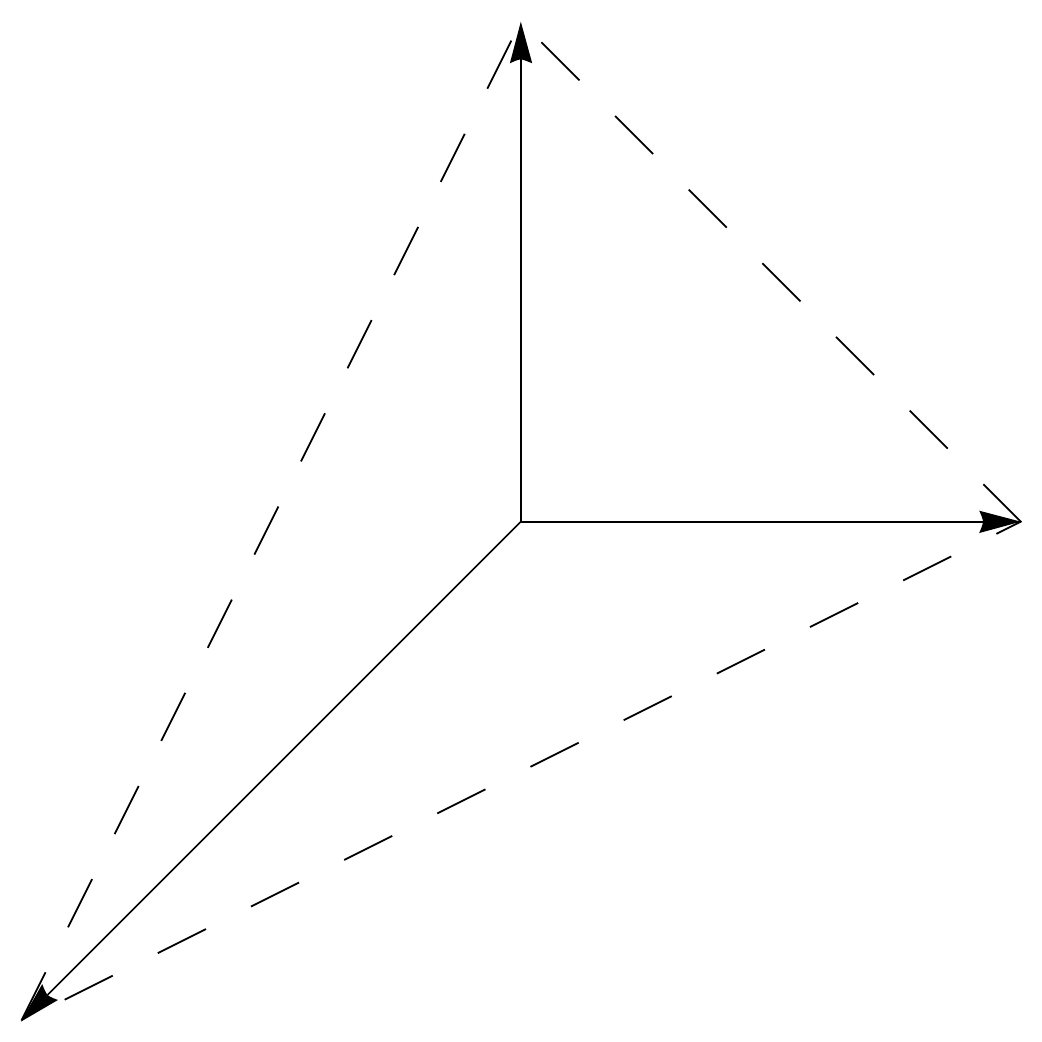}}}
\put(-120,-40){\makebox(0,0){ $B =\P^2$}}
\put(90,5){\makebox(0,0){
    \includegraphics[width=8cm]{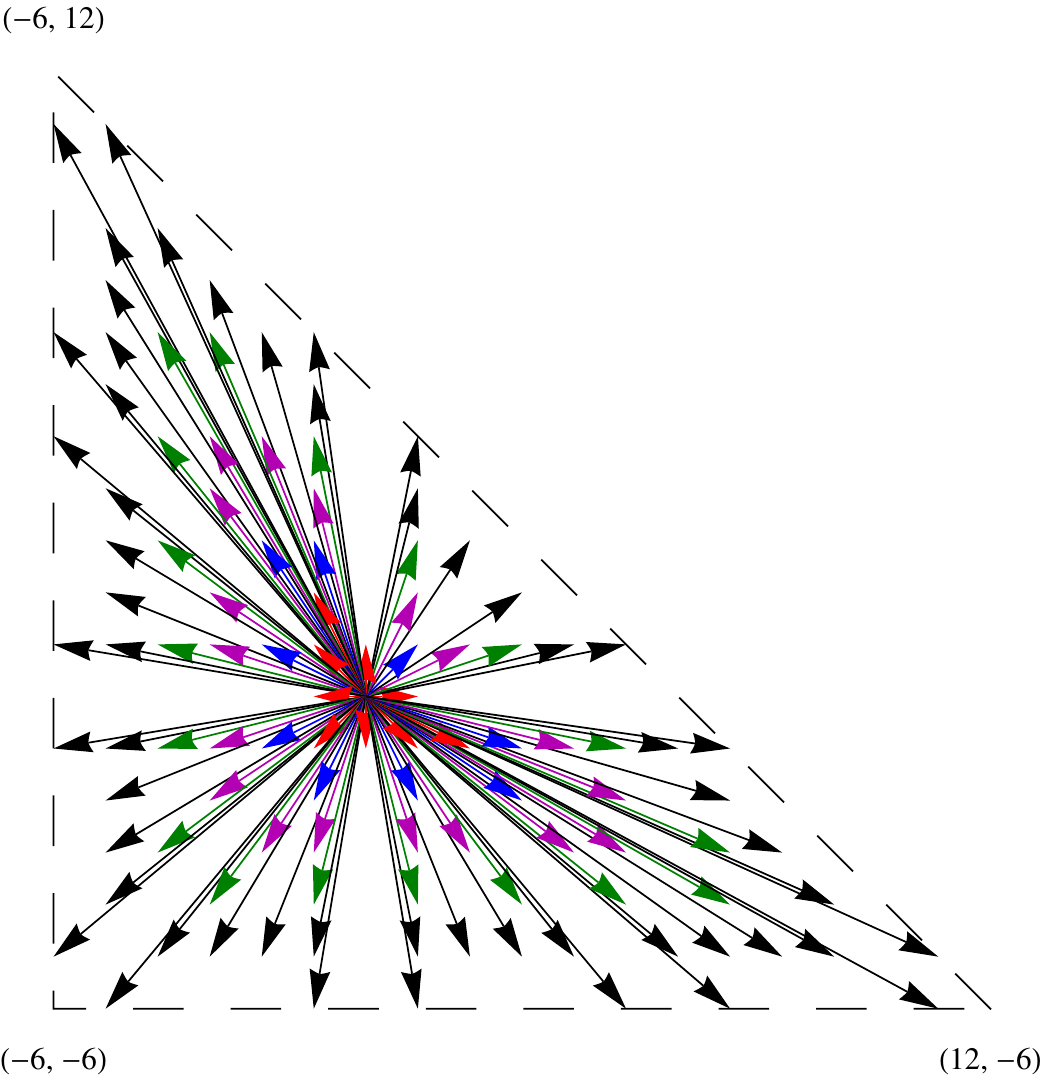}}}
\put(75,-120){\makebox(0,0){ $\tilde{B} \sim -6K_B$}}
\end{picture}
\caption[x]{\footnotesize The base $B = \P^2$ and the base $\tilde{B}$
of the
  mirror of the generic elliptic fibration over $B$, shown as toric varieties.
Rays that are red, blue, purple, green  correspond to
curves in the toric base $\tilde{B}$ that carry non-Higgsable $E_8,
F_4, G_2$, and $SU(2)$ gauge groups respectively.
}
\label{f:example-p2}
\end{figure}

As a simple first case, we consider the case of the base $B = \P^2$,
with a generic elliptic fiber given by an anti-canonical curve in the
toric fiber space $\P^{2, 3, 1}$ (i.e. the toric fiber $F_{10}$).  In
this case the polytope $\nabla$ has vertices
\begin{equation}
\{v_i\} =\{(0, 0; 1, 0), (0, 0; 0, 1), (1, 0; -3, -2), (0, 1; -3, -2),
(-1, -1; -3, -2)\} \,.
\label{eq:p2-nabla}
\end{equation}
Here the fiber is given by the slice with vertices $(0, 0; 1, 0), (0, 0; 0, 1), (0, 0;
-3, -2)$, and the projection onto the base projects the last two
coordinates to 0.  Since $B$ contains no curves of self-intersection
$-n$ with $n$ not a divisor of 12, $\nabla$ is immediately reflexive.
The dual polytope $\Delta$ is easily seen to have
vertices (see e.g. \cite{Huang-Taylor-1} for explicit analysis)
\begin{equation}
\{m_i\} =\{(0, 0; -1, 2), (0, 0; 1, -1), (-6, 12; -1, -1), (12, -6; -1, -1),
(-6, -6; -1, -1)\} \,.
\label{eq:example-w-P2}
\end{equation}
Under a coordinate transformation on the last two coordinates this
takes the form
\begin{equation}
\{w_i\} =\{(0, 0; 1, 0), (0, 0; 0, 1), (-6, 12; -3, -2), (12, -6; -3, -2),
(-6, -6; -3, -2)\} \,.
\label{eq:example-P2}
\end{equation}
This dual polytope $\Delta$ is again fibered by an $F_{10}$ fiber in
the $(0, 0; \cdot, \cdot)$ plane, and the projection onto the base
gives a base $\tilde{B}$ that has a toric description using the
primitive rays in the 2D polytope with vertices $(-6, 12),(12,
  -6),(-6, -6)$ (See Figure~\ref{f:example-p2}). The dual base polytope consists of the points in
$\Delta$ that can be associated with sections of the line bundle
  $-6K_B$.
The mirror polytopes $\nabla, \Delta$ both appear in the Kreuzer-Skarke
database, and give rise to elliptic toric hypersurface Calabi-Yau
threefolds with hodge numbers (2, 272) and (272, 2) respectively (see footnote \ref{ft1}).

We can characterize the mirror base $\tilde{B}$ as a toric variety by
the sequence of self-intersections of the toric rays calculated by equation (\ref{eq:dual-base})
\begin{equation}
\tilde{B} \rightarrow [[-11//-12//-12//-11//-12//-12//-11//-12//-12//]]
 \label{eq:p2-dual-base}
\end{equation}
where the notation $//$ denotes the sequence of self-intersections
\begin{equation}
//= -1, -2, -2, -3, -1, -5, -1, -3, -2, -2, -1\,.
\end{equation}
  This sequence of
self-intersections is a familiar sequence that connects $-12$ curves
that support $E_8$ (Kodaira type $II^*$) singularities in the elliptic
fibration (see e.g. \cite{clusters, Heckman-Morrison-Vafa}).
This sequence of self-intersections characterizes a face of the base
at distance 6 from the origin; similar structure for faces at
different distances from the origin are described in Appendix~\ref{sec:chains}.

Using methodology motivated by F-theory we can compute the Hodge
numbers of the generic elliptic fibrations over $B, \tilde{B}$
directly from the geometry of the bases \cite{Morrison-Vafa, Hodge}.  For $B$, all curves have
self-intersection above $-3$, so there is no non-Higgsable gauge
group.  From the Shioda-Tate-Wazir formula we have
\begin{equation}
 h^{1,1} (X) = h^{1,1} (B) +1 = 2 \,.
\end{equation}
From the gravitational anomaly condition, we have
\begin{equation}
 h^{2,1} (X) = 273-29 (h^{1,1} (B) -1) - 1 = 272 \,.
\end{equation}
For the mirror the computation is a bit more complicated.  On each
$-11$ curve there is a single (4, 6) point that must be blown up so
that the total space has a smooth Calabi-Yau resolution
\cite{clusters}.  Before these blowups the number of toric curves is
108, so $h^{1,1} (B) = 108 - 2+ 3=  109$.  The non-Higgsable gauge group
from the curves of negative self-intersection below $- 1$ is $G =
E_8^9 \times F_4^9 \times (G_2 \times SU(2))^{18}$, with rank 162.  We
then have
\begin{equation}
  h^{1,1} (\tilde{X}) = h^{1,1} (\tilde{B}) + {\rm rank}\ G +1
  = 109+162+1 = 272 \,.
\end{equation}
On the other hand, each $G_2 \times SU(2)$ non-Higgsable factor is
associated with 8 charged matter hypermultiplets, so we have
\begin{equation}
  h^{2,1} (\tilde{X}) =
  273-29 (h^{1,1} (B) -1)+ {\rm dim}\ G - m_{\rm NH}-1
  = 273-29 (108) +9 (248+52+34) -144-1 = 2 \,.
\end{equation}

\subsection{Example: self-mirror Calabi-Yau threefold with Hodge
  numbers (251, 251)}

\begin{figure}
\centering
\begin{picture}(200,100)(- 100,- 50)
\put(0,0){\makebox(0,0){ \includegraphics[width=14cm]{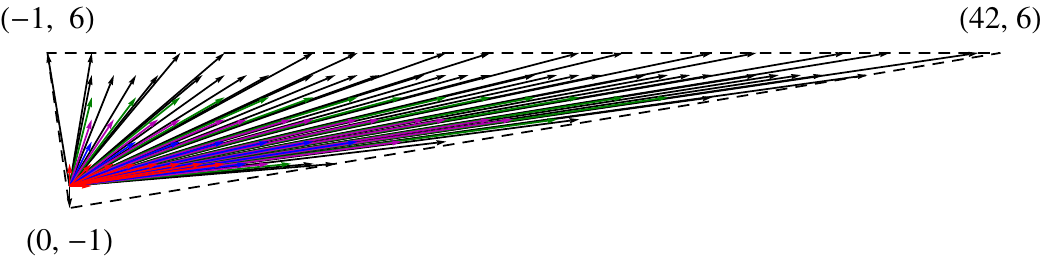}}}
\put(0,-30){\makebox(0,0){ $B = \tilde{B}$}}
\end{picture}
\caption[x]{\footnotesize The base $B = \tilde{B}$ over which the
  generic elliptic fibration is the self-mirror Calabi-Yau threefold
  with Hodge numbers $(251, 251)$.
Note that the inner product between each pair of vertices is $-6$, so
that $B = \tilde{B} \sim -6K_B$.
Rays that are red, blue, purple, green  correspond to
curves in the toric base $\tilde{B}$ that carry non-Higgsable $E_8,
F_4, G_2$, and $SU(2)$ gauge groups respectively.
}
\label{f:251}
\end{figure}

We now consider the self-mirror Calabi-Yau threefold with Hodge
numbers (251, 251) at the central peak of the ``Hodge shield''.  This
polytope can be put into a coordinate system where the vertices are
\begin{equation}
\{v_i\} =\{(0, 0; 1, 0), (0, 0; 0, 1), (-1, 6; -3, -2), (0, -1; -3, -2),
(42, 6; -3, -2)\} \,.
\end{equation}
This polytope is self-dual, $\nabla = \Delta$, up to a coordinate transformation, and is clearly a $\P^{2,
  3, 1}$ fibration over the base $B$
with vertices $(-1, 6), (0, -1), (42, 6)$.  
Since the inner product between each pair of these vertices is $-6$,
the procedure of constructing the mirror base $\tilde{B}$ from
monomials in ${\cal O} (-6K_B)$ gives $B = \tilde{B}$.
The primitive rays in this
base (Figure~\ref{f:251}) give a toric surface with a sequence of
self-intersections
\begin{equation}
[[0,  6,
-12//-11//-12//-12//-12//-12//-12//-12//-12]] \,.
\end{equation}

Again using the relationship between geometry and F-theory physics we
can compute the Hodge numbers directly from the geometry of the base.
There are 99 toric curves in the base, with one $-11$ curve that must
be blown up, so $h^{1, 1}(B) = 99-2 + 1 = 98$.  The non-Higgsable
gauge group is $G = E_8^9 \times F_4^8 \times (G_2 \times
SU(2))^{16}$, with rank 152, so
\begin{equation}
h^{1, 1}(X) = h^{1, 1} (B) + {\rm rk}\, G + 1 =  251\,.
\end{equation}
Similarly,
\begin{equation}
  h^{2,1} (\tilde{X}) =
  273-29 (h^{1,1} (B) -1)+ {\rm dim}\ G - m_{\rm NH}-1
=251 \,.
\end{equation}
 
\subsection{Example: generic elliptic fibration over $\F_n$ (Hodge
  numbers (3, 243), \ldots, (11, 491)}
\label{sec:Hirzebruch}

As further examples we consider the generic elliptic fibrations over
the Hirzebruch surfaces $B =\F_n$.  In each case the mirror Calabi-Yau is
elliptic over a base constructed from $-6K_B$, though in some cases
the mirror is not a generic elliptic fibration but has some tuning.
We describe several specific cases in detail and summarize the
complete set for all $n = 0, \ldots, 12$ in
Table~\ref{f:Hirzebruch-bases} and Table \ref{f:Hirzebruch-bases-nontoric}.  The Hirzebruch surface $\F_n$ can be
described by the toric rays $(0, 1), (1, 0), (0, -1),(-1, -n)$.  For
$n = 0, 1$ these are the vertices of the associated 2D polytope, while
for $n > 1$ $(0, -1)$ is not a vertex and the other three are.  The
dual base $\tilde{B} \sim -6K_B$ is thus
characterized by the toric variety with a fan given by all primitive
rays in the polytope defined by the vertices
\begin{equation}
\{(-6, -6),(-6, 12/n), (6 (n + 1), -6)\} \,,
\label{eq:Hirzebruch-dual}
\end{equation}
when $n  \geq 2$, 
$\{(-6, -6), (-6, 12), (6, 0), (6, -6))\}$ for $n = 1$,
and $\{(-6, -6), (-6, 6), (6, 6), (6, -6)\}$ for $n
= 0$.

When $n
| 12, n > 1$,  the
vertices of the polytope (\ref{eq:Hirzebruch-dual}) containing the
dual base are integral; from this it follows that the
 polytope $\nabla$ is reflexive, since $\ds$ has the form of
 (\ref{eq:example-w-P2}),  where the last set of vertices are given by
 $(m^{(I)}; -1, -1)$, with $m^{(I)}$ vertices of the polytope (\ref{eq:Hirzebruch-dual}).
The polytope $\nabla$  has vertices\cite{Huang-Taylor-1}
\begin{equation}
\{v_i\} =\{(0, 0; 1, 0), (0, 0; 0, 1), (1, 0; -3, -2), (0, 1; -3, -2),
(-1, -n; -3, -2)\} \,.
\label{eq:nabla-Hirzebruch}
\end{equation}
As in the preceding examples we can read off the sequence of
self-intersections of the toric rays associated with primitive rays in
(\ref{eq:Hirzebruch-dual}).  We give a few explicit examples.

{\bf $\F_{12}$}:\\
The polytope $\nabla$ is reflexive, and the Hodge numbers of the
associated Calabi-Yau threefold are $(11, 491)$; this is the largest
possible value of $h^{2, 1}(X)$ for any elliptic Calabi-Yau threefold \cite{Hodge}.
In this case the vertex $(-6, 1)$ of $\tilde{B}$ is
primitive and corresponds to a curve of self-intersection 0, and the
vertices $( -6, -6)$ and $(78, -6)$ each go to primitive rays $(-1,
-1)$ and $ (13, -1)$ associated with curves of self-intersection 11.
The sequence of self-intersections of the toric rays for $\tilde{B}$
is then
\begin{equation}
[[0, -12//-11(//-12)^{12}//-11//-12]]
\end{equation}
Blowing up the base at two points on the $-11$ curves we can confirm
that the Hodge numbers of the generic elliptic fibration,
corresponding to the polytope $\Delta$, are (491, 11) as expected.

\begin{figure}
\begin{center}
 \includegraphics[width=8cm]{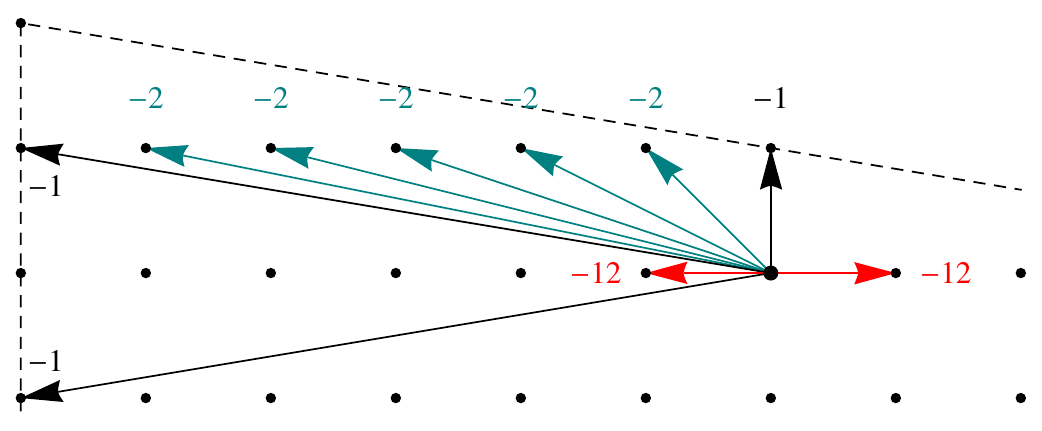}
\end{center}
\caption[x]{\footnotesize  Part of the ray structure of the toric base
$\tilde{B}$ for the mirror $\tilde{X}$ of the generic elliptic
  fibration over the Hirzebruch surface $\F_6$.}
\label{f:f6-top}
\end{figure}

{\bf $\F_6$}:\\
 In this case again $\nabla$ defined through
(\ref{eq:nabla-Hirzebruch}) is immediately reflexive.  
The Hodge numbers for the associated Calabi-Yau threefold are (9,
321), which can be immediately determined from the non-Higgsable $E_6$
gauge group over the $-6$ curve in $B$.
In this case,
however, the vertex $(-6, 2)$ is not primitive.  The top part of the
toric diagram for $\tilde{B}$ is shown in
Figure~\ref{f:f6-top}, and contains the sequence of curves of self-intersection $-1,
-2, -2, -2, -2, -2, -1$.  Unlike in the case of $\F_{12}$ the mirror
polytope $\Delta$ is not a generic elliptic fibration over
$\tilde{B}$.  And indeed, the generic elliptic fibration over $\tilde{B}$ has Hodge
numbers $(317, 17)$ rather than the values of $(321, 9)$ expected from
mirror symmetry.
This can be understood from the fact that the vertex $(-6, 2)$
is not present in $\tilde{B}$, so the monomials in $-6K_{\tilde{B}}$
are not simply the points in $B =  \F_{6}$ but also include the
lattice points $(0, -k), k \in\{4, 5, 6\}$.
In the polytope $\nabla$
these monomials are all set to vanish.  Computing the resulting gauge
group structure on $\tilde{B}$ we find that there is a gauge group
$SU(2) \times G_2 \times SU(2)$ tuned on
the middle sequence of three $-2$ curves.  This gives an additional
rank contribution of 4 to $h^{1, 1}(\tilde{X})$  from the gauge group
and there are 
contributions to $h^{2, 1}(\tilde{X})$ of $+ 20$ from the dimension of
the tuned gauge group, and $-28$ from the charged matter fields,\footnote{Note that the correct counting of
  $h^{2,1}$  considers only the matter fields   charged under the
  Cartan subalgebra. The counting in this case is equivalent to that of the
  rank preserving tuning of $SU(2) \times SU(3) \times SU(2)$, where the $SU(3)$ has 6 hypermultiplets charged
in the fundamental ({\bf 3})
representation, 4 of which are in bifundamentals
with the $SU(2)$ factors.  In this case the $SU(3)$ fundamentals
combine in pairs into $G_2$ fundamentals, and
the $G_2$ has one additional fundamental ({\bf 7}),
which contributes another 6 hypermultiplets charged under the Cartan, canceling the difference in dimension between $SU(2)$ and $G_2$. This rank-preserving tuning between $SU(2)$ and $G_2$  connects two phases of the same Calabi-Yau geometry.}
so that the correct Hodge numbers are found for the tuned Calabi-Yau
\begin{equation}
(321, 9) = (317, 17) + (4, -8) \,.
\end{equation}

When $n$ does not divide $12$, there are additional vertices of
$\nabla$ that must be included to attain a reflexive polytope from the
simple stacking of $\F_n$.

{\bf $\F_5$}:\\
For example, when $n=5$,
$(0,-3,-3,-2)$ 
is
an additional vertex
that must be included with the set in (\ref{eq:nabla-Hirzebruch}) for
$\nabla$ to be reflexive\footnote{ Additional vertices from tops for
  all generic fibrations over Hirzebruch surfaces can be looked up in
  Table 11 in \cite{Huang-Taylor-1}.}, so the projection of the whole
$\nabla$ polytope to the base plane is instead the polytope defined by
the vertices
\begin{equation}
\{(1,0), (0,1), (-1,-5), (0,-3)\},
\label{f5}
\end{equation}
and the convex hull of the toric fan of $B$ lies within this polytope. On the dual
side, $\tilde X$ is a generic elliptic fibration with Hodge numbers
$(295,7)$. The polytope defined by all monomials in
$\mathcal{O}(-6K_B)$ (defined using the rays in $B$ from equation
(\ref{f5})) has the set of vertices
\begin{equation}
\{(-6,2),(-4,2),(-6,-6),(36,-6))\}.
\end{equation}
This is the projected polytope of the whole $\Delta$ to the base. 
 The primitive rays in this projected polytope define the base
 $\tilde{B}$, which is characterized by the self-intersection sequence
\begin{equation}
[[-12//-11//(-12//)^6-11//-12,-1,-2,-2,-3,-1,-3,-2,-2,-1]],
\label{eq:f5-example}
\end{equation}
and the vertices of the convex hull of the toric fan of $\tilde B$ are
\begin{equation}
\{(-5, 2), (-6, 1), (1, 1), (-6, -5), (-5, -6), (31,-5), (35, -6)\},
\end{equation}
which is contained in the projected polytope.
 In this case the mirror $\Delta$ is again a generic (non-tuned)
 elliptic fibration over the mirror base $\tilde{B}$, and the Hodge
 numbers can be computed directly from the non-Higgsable clusters on
 the base with intersections (\ref{eq:f5-example}).

We list the results of the remaining $\F_n$ cases in Table
\ref{f:Hirzebruch-bases} for $0\leq n \leq 8, n=12$.\footnote{Thanks
  to Yinan Wang for a computation showing that tunings were missing in
the $\F_3, \F_4$ examples in the original version of the paper.}
For $n=9,10,11$,
additional blowups at points in the base that may be toric or
non-toric are required to
support a flat elliptic fibration, and there are various ways to
resolve these bases (see Table 15 in \cite{Huang-Taylor-1}). We find
that the corresponding mirror fibrations are generic models over
different dual bases $\tilde B$ for different resolutions. The results
are listed in Table \ref{f:Hirzebruch-bases-nontoric}.

\begin{landscape}
\begin{table}[]
\begin{tabular}{|c|c|l|l|}
\hline
$B$       & $(h^{1,1},h^{2,1})$ & Mirror base $\tilde B$                                                     & tuning                              \\ \hline
$\F_0$    & (3,243)             & $(-12//-11//)^3-12//-11,-1,-2,-2,-3,-1,-5,-1,-3,-2,-2,-1$                  &                                     \\ \hline
$\F_1$    & (3,243)             & $-12//-11//(-12//)^2-11//-12//-11//-11,-1,-2,-2,-3,-1,-5,-1,-3,-2,-2,-1$   &                                     \\ \hline
$\F_2$    & (3,243)             & $-12//-11//(-12//)^3-11//-12//-10,-1,-2,-2,-3,-1,-5,-1,-3,-2,-2,-1$        &                                     \\ \hline
$\F_3$    & (5,251)             &
$-12//-11//(-12//)^4-11//-12,-1,-2,-2,-3,-1,-5,-1,\mathbf{-3},-1,-5,-1,-3,-2,-2,-1$
& $\gg_2$                                     \\ \hline
$\F_4$    & (7,271)             &
$-12//-11//(-12//)^5-11//-12,-1,-2,-2,-3,-1,
\mathbf{-4},-1,-3,-2,-2,-1$             & 
$\gf_4$                                    \\ \hline
$\F_5$    & (7,295)             & $-12//-11//(-12//)^6-11//-12,-1,-2,-2,-3,-1,-3,-2,-2,-1$                   &                                     \\ \hline
$\F_6$    & (9,321)             & $-12//-11//(-12//)^7-11//-12,-1,-2,\mathbf{-2,-2,-2},-2,-1$                & $\gsu(2)\oplus \gg_2\oplus \gsu(2)$ \\ \hline
$\F_7$    & (10,348)            & $-12//-11//(-12//)^8-11//-12,-1,-2,\mathbf{-2,-2},-2,-1$                   & $\gsp(1)\oplus\gsp(1)$              \\ \hline
$\F_8$    & (10,376)            & $-12//-11//(-12//)^9-11//-12,-1,-2,\mathbf{-2,}-2,-1$                      & $\gsp(1)$                           \\ \hline
$\F_{12}$ & (11,491)            & $-12//-11//(-12//)^{13}-11//-12,0$                                         &                                     \\ \hline
\end{tabular}
\caption[x]{\footnotesize For each Hirzebruch base $B = \F_n$, $0\leq n \leq 8, n=12$ the
  Hodge numbers of the generic elliptic fibration $X$ over $B$, the toric
  structure of the base $\tilde{B}$ of the mirror Calabi-Yau threefold
$\tilde{X}$, and any gauge algebra tuned (on the boldfaced curves) in
  $\tilde{B}$. Note that the tunings on the mirror bases for $\F_3,
\F_4$ are rank-preserving tunings that do not change the Hodge numbers
(corresponding to a different phase of the same Calabi-Yau); the
generic elliptic CY fibrations over the mirror bases in these cases
are mirror to corresponding rank-preserving tunings on the $-3, -4$
curves of the original $\F_3, \F_4$ respectively.}
\label{f:Hirzebruch-bases}
\end{table}

\end{landscape}

\begin{table}[]
\begin{tabular}{|c|c|c|l|}
\hline
$B$                        & ($h^{1,1},h^{2,1}$)         & Resolved $B$ & Mirror base $\tilde B$                       \\ \hline
\multirow{6}{*}{$\F_9$}    & \multirow{6}{*}{$(14,404)$} &  \includegraphics[width=1.4cm]{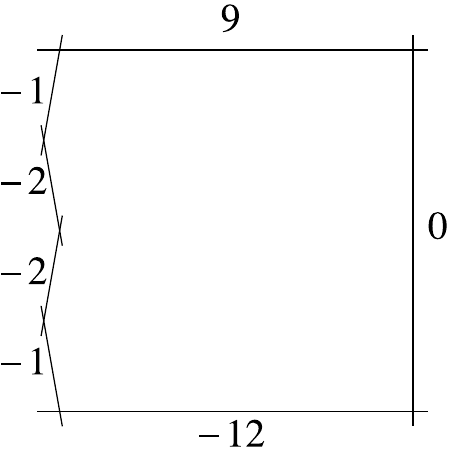}      & $-12//-11//(-12//)^{10}-11//-9,0$            \\ \cline{3-4} 
                           &                             &
\includegraphics[width=1.4cm]{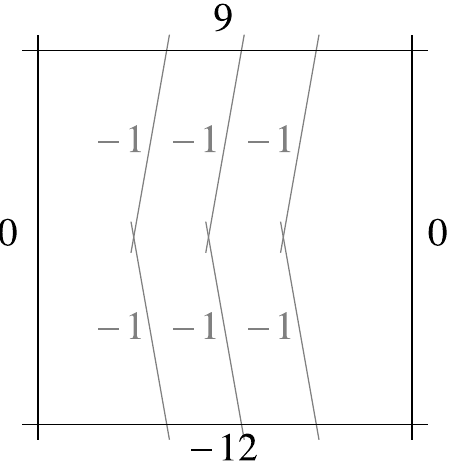} (*)      & $-12//-11//(-12//)^{10}-11//-12,-1,-2,-2,-1$ \\ \cline{3-4} 
                           &                             &  \includegraphics[width=1.4cm]{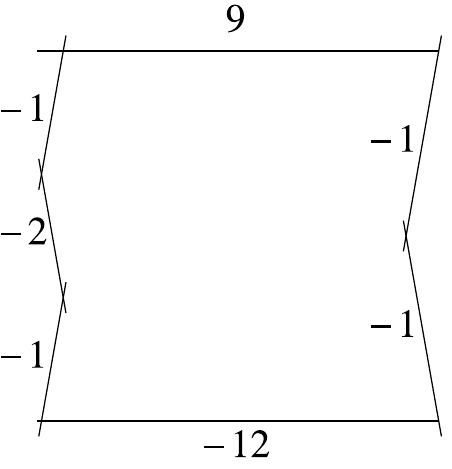}      & $-11//-11//(-12//)^{10}-11//-10,0$           \\ \cline{3-4} 
                           &                             &  \includegraphics[width=1.4cm]{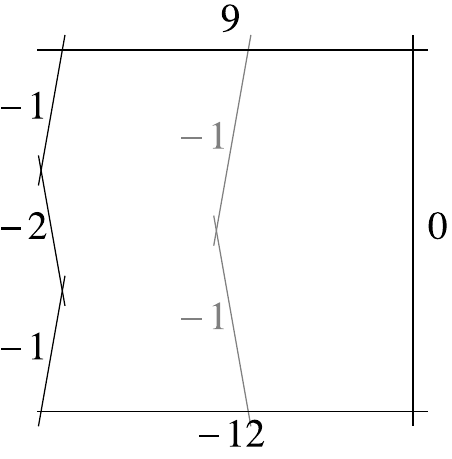}      & $-12//-11//(-12//)^{10}-11//-10,-1,-1$       \\ \cline{3-4} 
                           &                             &  \includegraphics[width=1.4cm]{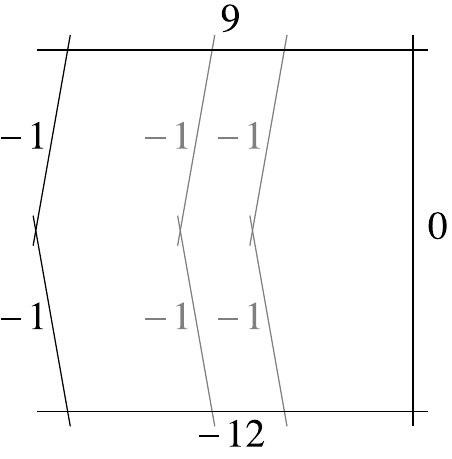}      & $-12//-11//(-12//)^{10}-11//-11,-1,-2,-1$    \\ \cline{3-4} 
                           &                             & \includegraphics[width=1.4cm]{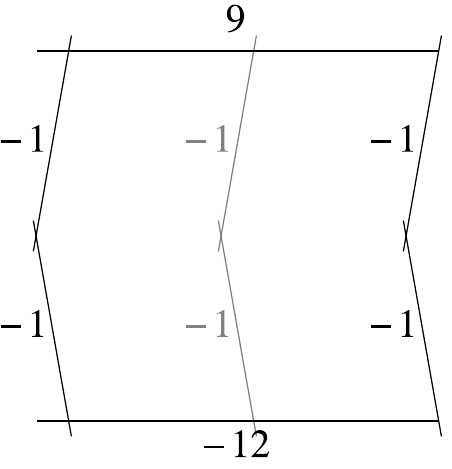}      & $-11//-11//(-12//)^{10}-11//-11,-1,-1$       \\ \hline
\multirow{4}{*}{$\F_{10}$} & \multirow{4}{*}{$(13,433)$} &
\includegraphics[width=1.4cm]{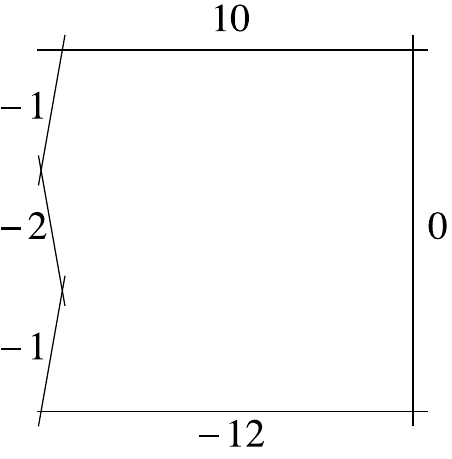} (*)     & $-12//-11//(-12//)^{11}-11//-10,0$           \\ \cline{3-4} 
                           &                             & \includegraphics[width=1.4cm]{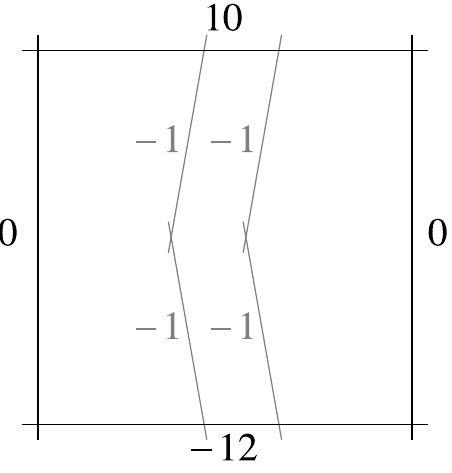}     & $-12//-11//(-12//)^{11}-11//-12,-1,-2,-1$    \\ \cline{3-4} 
                           &                             & \includegraphics[width=1.4cm]{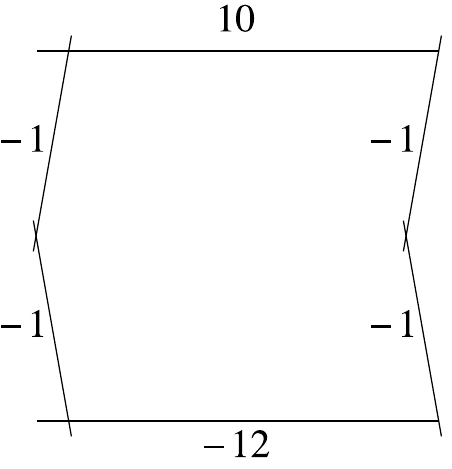}     & $-11//-11//(-12//)^{11}-11//-11,0$           \\ \cline{3-4} 
                           &                             & \includegraphics[width=1.4cm]{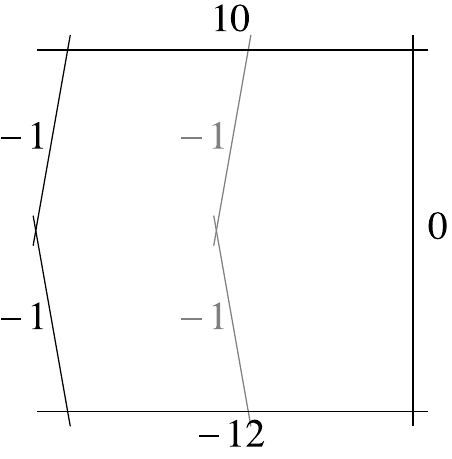}     & $-12//-11//(-12//)^{11}-11//-11,-1,-1$       \\ \hline
\multirow{2}{*}{$\F_{11}$} & \multirow{2}{*}{$(12,462)$} & \includegraphics[width=1.4cm]{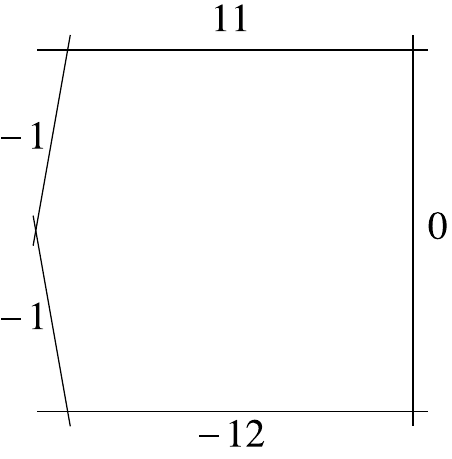}     & $-12//-11//(-12//)^{12}-11//-11,0$           \\ \cline{3-4} 
                           &                             &
\includegraphics[width=1.4cm]{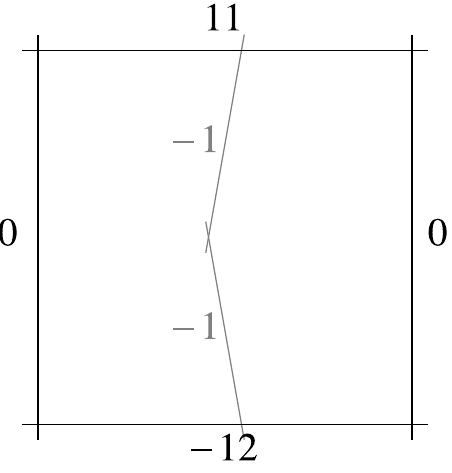} (*)     & $-12//-11//(-12//)^{12}-11//-12,-1,-1$       \\ \hline
\end{tabular}
\caption[x]{\footnotesize For each Hirzebruch base $B = \F_n$, $n=9, 10, 11$ the
  Hodge numbers
 of the generic elliptic fibration $X$ over
 resolved $B$, the toric
  structure of the resolved base and the base $\tilde{B}$ of the mirror Calabi-Yau threefold
$\tilde{X}$.
The bases marked with (*) are resolved at generic non-toric points,
and correspond to the naive stacking models with the given Hirzebruch base,
while the other bases are resolved at combinations of toric and
non-toric points.
}
\label{f:Hirzebruch-bases-nontoric}
\end{table}

\section{Some further examples}
\label{sec:further-examples}

In this section we consider some further examples that go beyond
generic CY elliptic fibrations over toric bases.  We first consider 
some
cases of Tate tunings of the generic elliptic fibrations.  As
discussed in more detail in \cite{Huang-Taylor-1}, a Tate tuning over
a toric base generally gives a reflexive 4D polytope with a standard
stacking form and the $F_{10}$ fiber, where the Tate tuning
corresponds to reducing the monomials in $\Delta$.  There is a natural
correspondence between these Tate tunings and tops involving
additional lattice points in $\nabla$.  
In general, for a Tate tuning
where $a_1, \ldots, a_6$ vanish to orders $[a_n]_i$ on the toric
divisor associated with the ray $v^{B}_i$, 
the vanishing of the monomials in  $a_n$ corresponds to a removal of
the points in $\Delta$ that lie over the point in $\ds_2$ associated
with sections of the line bundle ${\cal O} (-nK_B)$.  For many Tate
tunings, as for the generic elliptic fibration, the remaining
points in the toric representation of ${\cal O} (-6K_B)$
are a superset of the remaining points in the other ${\cal O}
(-nK_B)$'s.  This is equivalent to the condition that the mirror
polytope $\Delta$ has a standard stacking form so that all rays in the
base have a preimage under the fiber projection of the form $(\cdot,
\cdot; -1, -1)$.  In such situations,
the set of rays
of the dual base becomes
\begin{eqnarray}
\{w_i^{\tilde B}\}= V^B (a_6)\,,
\label{eq:dual-base-tuned}
\end{eqnarray}
where\footnote{Note that $V^B (a_n)$ can also be described simply as a set of 
primitive rays in 2D toric coordinates
associated with the monomials in $a_n$, with the product $\Pi_i z_i$
of the toric variables $z_i$
associated with the rays of $\Sigma_B$ taken as the origin.
}
\begin{equation}
V^B  (a_n)=
\{ w =(w_1,w_2)\rvert \text{GCD}(w_1, w_2)=1, 
w \cdot v_i^{B} \geq  - n + [a_n]_i \;\;\forall v^B_i \in \Sigma_B\}\,.
\label{eq:V}
\end{equation}
We describe explicit examples of this 
in \S\ref{sec:tuning} and  \S\ref{sec:tuning-2}.
More generally, there are some Tate tunings (such as $SU(6)$) that
place such stronger constraints on $a_6$ than on other coefficients
$a_n$.  This can occur when $6-[a_6]_i < n-[a_n]_i$ for some $i$
and $n
  \leq 4$.
In such cases, the mirror $\Delta$ is no longer a standard
stacking but we can still give an explicit description of the base
$\tilde{B}$ in terms of $B$ and the Tate tuning,
\begin{equation}
\{w_i^{\tilde{B}}\} = \cup_{n  \in\{1, 2, 3, 4, 6\}}V^B  (a_n)\,.
\label{eq:dual-base-general}
\end{equation}
This is essentially a rewriting of (\ref{eq:base}) for the mirror
polytope $\Delta$ when $\dd$ is a standard stacking type polytope with
the fiber $F_{10}$.  We describe an explicit example of this 
in \S\ref{sec:tuning-3}. 
In terms of the monomial polytope $\Delta$,
the set of points determined by
 equation (\ref{eq:V}) are given by
\begin{eqnarray}
V^B(a_n) = 
\{ (m_1,m_2)/\text{GCD}(m_1, m_2)\rvert  
 m=(m_1,m_2, m^{\tilde F}(n)) \in \Delta  \}\,,
 \label{eq:dual-base-tuned-2}
\end{eqnarray}
where $m^{\tilde F}(n)$ represents the two coordinates of the
particular lattice point $m^{(II)}\in \Delta_2$ that is associated
with monomials in  $a_n$. 
This formulation was used in some of the following explicit computations.
In the coordinate system we adopt in this paper
\begin{eqnarray}
\label{45}
&&m^{\tilde F}(n=1,2,3,4,6)=\{(0,0),(-1,1),(0,-1),(-1,0),(-1,-1)\},\\
\label{46}
&&m^F (n=1,2,3,4,6)=\{(0,0),(-1,0),(-1,-1),(-2,-1),(-3,-2)\}.
\end{eqnarray} For example, 
$V^B  (a_6)=
\{ (m_1,m_2)/\text{GCD}(m_1, m_2)\rvert  
 m \in \Delta \text{ of the form } (m_1,m_2,-1,-1)\}\,.$ In principle, we can use this same kind of
analysis to describe other kinds of fibrations with general fiber
types $F_i$, though details of the structure of tuned models will be
different and depend on tops for the different fibers
\cite{Bouchard-Skarke}.  We leave a more complete analysis of the
general situation to future work.

In \S\ref{sec:other-fiber}, we consider a case of a stacked fibration
with another fiber type, in \S\ref{sec:non-fibered} we illustrate a
case of a fibered polytope where the mirror is not fibered,
and in \S\ref{sec:4D} we give an example of the factorization of
mirror symmetry for a Calabi-Yau fourfold constructed as a generic
elliptic fibration over a toric threefold base.  These examples simply
illustrate some of the directions in which the framework developed
here can be extended, and a more detailed analysis of these directions
is left for the future.

\subsection{Tunings of generic fibrations (Example:
tuning an $SU(2)$ on $\P^2$)}
 \label{sec:tuning}

We consider tuning an $SU(2)$ on one of the $+ 1$-curves in the $\P^2$ base
of the generic fibration model in \S\ref{sec:example-p2}. The
corresponding $\Delta$ polytope of the tuned model can be constructed by
reducing the set of lattice points in the $\Delta$ polytope of the generic model (detailed examples
of constructing tuned polytope models can be found in Appendix A in
\cite{Huang-Taylor-1}); this corresponds to the polytope in the case database with data: M:316 7 N:11 6 H:3,231. The new Hodge numbers
match with the prediction from F-theory physics of tuning the $SU(2)$
on a  $+ 1$-curve $(2,272)+(1,-41)=(3,231).$ The standard stacking polytope
$\nabla$ has vertices
\begin{eqnarray}
\nonumber
\{v_i\} =\{ (0, 1; -3, -2), (-1, -1; -3, -2), (1, 0; -3, -2), (1, 
  0; -2, -1), (0, 0; 1, 0), (0, 0; 0, 1)\}.
\end{eqnarray}
Now there is a non-trivial top over the base divisor associated with
the ray $(1,0)$ due to the $SU(2)$ tuning over the divisor, which we
can interpret in terms of a Tate tuning in $\ds$. $\nabla$ has the
usual
$(0, 0; v^F)$ form of $F_{10}$ fiber as the only 2D subpolytope $\nabla_2$. The dual polytope $\Delta$ has vertices
\begin{eqnarray}
\nonumber
\{w_i\}=\{(-4,-6;-1,-1),(-4,10;-1,-1),(-2,-2;-1,1),&&(-2,4;-1,1),(12,-6;-1,-1),\\
&&(0,0;-1,2),(0,0;1,-1)
\}.
\label{eq:p2-tuning-dual}
\end{eqnarray}
The $\Delta$ polytope has multiple distinct fibrations, 
consisting of four $F_{10}$ fibers, three $F_{13}$ fibers,
and one $F_{16}$ fiber.  In particular, there is a fiber $\Delta_2$
dual to $\nabla_2$, which has lattice points also in the form $(0,
0;w^F)$. 
This gives the mirror fibration of the $\Delta$ polytope
$\tilde X$ when viewed as a fan polytope.\footnote{Note that $\Delta$
  does not project onto the other fibers (the projection strictly
  contains the other fibers); otherwise $\nabla$ would have had more
  than one fiber subpolytope.}  
We can calculate the dual base $\tilde B$ by
equation
(\ref{eq:base}) for the polytope $\Delta$, 
giving the self-intersection sequence
\begin{eqnarray}
  \tilde B \rightarrow &&[[-12//-12//-11// -12// -12, -1, -2, -2, -3, -1, -5, -1, -3, -2, \\
      & & -1, -8, -1, -2, -3, -2, -1, 
      -8, -1, -2, -3, -2, -1, -8, -1, -2, -3, -2,\nonumber \\
      && -1, -8, -1, -2, -3, -1, -5, -1, -3, -2, -2, -1]].
\nonumber
\end{eqnarray}
This is
exactly the base over which the generic elliptic fibration has Hodge numbers
$(231, 3)$. Therefore, we know $\Delta$ is the fan polytope associated
with this generic fibration model.  This generic fibration is thus
the mirror fibration to the tuned $SU(2)$ fibration model.

Note that this base has a smaller toric diagram than the base
described in (\ref{eq:p2-dual-base}).  There are additional
constraints on the rays $w$ in (\ref{eq:dual-base}) in this case
coming from base
rays $v^B$ over points in the fiber $\dd_2$ other than $v_s$.  
We can see how this works explicitly as an example of
(\ref{eq:dual-base-tuned}).  Tuning an SU(2) gauge factor using a Tate
tuning over the divisor in $\P^2$ associated with $v_3^B = (1, 0)$
requires tuning the coefficients in (\ref{eq:Tate}) to vanish to
orders $([a_1], [a_2], [a_3], [a_4], [a_6]) = (0, 0, 1, 1, 2)$
on this divisor.
Thus,  the rays in $\tilde{B}$ are
restricted to a subset of those in ${\cal O} (-6K_B)$, which satisfy
the additional condition $w \cdot v_3^B \geq -4$, so $w_1 \geq -4$.
This is already no stronger than the constraint on any of the other
sections ${\cal O} (-nK_B), n \leq 4$, so we can use
 (\ref{eq:dual-base-tuned}) and there are no other contributions from
the more general formula (\ref{eq:dual-base-general}).
From Figure~\ref{f:example-p2} we see that the base $\tilde{B}$ is
thence associated with the primitive rays in the polytope with
vertices $(-4, 10), (-4, -6), (12, -6)$.  This matches perfectly with
the projection from (\ref{eq:p2-tuning-dual}).

As $\Delta$ is also a generic elliptic fibration associated with a
 standard stacking polytope with stacking point
$w_s=(-1,-1)$, we can calculate $B\sim\mathcal{O}(-6K_{\tilde B})$, 
and confirm that the monomials in $\mathcal{O}(-6K_{\tilde B})$ are
all the lattice points of the form $(\cdot,\cdot, -3, -2)$
in the original polytope $\nabla$.

It is interesting to note that in this case while there is a tuning of
the fibration on the $\dd$ side, corresponding to a reduction in the
size of $\tilde{B}$ on the $\ds$ side, the mirror is still a generic
elliptic fibration over the new base.  We now consider a case where
there are tunings on both sides.

\subsection{Tunings of generic fibrations over base $B$
and mirror base $\tilde{B}$}
 \label{sec:tuning-2}

We now consider an example where both the fibration and the mirror
fibration are tuned models. There is only one polytope associated with
a CY3 with Hodge numbers $(6, 248)$, which is a standard stacking
polytope. The polytope $\nabla$ has vertices
\begin{eqnarray}
\nonumber
\{v_i\}  &=
&\{(1,0;-3,-2),(0,1;-3,-2),(-1,-3;-3,-2),(0,-1;-1,0),(0,-2;-3,-2),\\
& & \hspace*{0.1in}(0,0;1,0),(0,0;0,1)\}.
\end{eqnarray}
The obvious $F_{10}$ fiber
in the plane $(0, 0; \cdot, \cdot)$ is the only fiber of $\nabla$. The  associated CY3 is a Tate tuned model over the base $\F_3$ with $\gso(7)$ gauge symmetry enhanced on the $-3$-curve. The dual polytope $\Delta$ has vertices
\begin{eqnarray}
\nonumber
\{w_i\}& = &\{(  24,  -6;  -1,  -1),
  ( 0,   2;  -1,  -1),
 ( -6,  -6;  -1,  -1),
 ( -6,   2;  -1,  -1),
 ( -2,   2;  -1,   0),\\
& & \hspace*{0.1in}
 ( -4,   2;  -1,   0)
  (  0,   0;   1,  -1),
  (  0,   0;  -1,   2)
\} \,.
\end{eqnarray}
The dual polytope $\Delta$ has three $F_{10}$ fibers and one $F_{13}$
fiber. The dual fiber with lattice points in the form $\{0, 0; \cdot, \cdot\}$
gives the mirror fibration with Hodge numbers $(248, 6)$.  $\Delta$ is
a standard stacking polytope with stacking point $w_s=(-1,-1)$ with
respect to the fiber. The mirror base, by direct calculation (taking the
rays of $\tilde B$ to be the primitive rays of all the projected 4D
lattice points in $\Delta$ using (\ref{eq:base})
on $\Delta$) is
\begin{eqnarray}
\tilde B \rightarrow &&[[-4,-1,-4,-1,-3,-2,-2,-1,\\
&&-12//-11//-12//-12//-12//-12//-11//-12,-1,-2,-2,-3,-1]].
\nonumber
\end{eqnarray}
The generic fibration over $\tilde B$ has Hodge numbers (247, 7), so this is a tuned model. By explicit analysis of the Weierstrass model over $\tilde B$ associated with the polytope $\Delta$, we know there are enhanced gauge symmetries $\gso(9)\oplus\gsp(1)\oplus\gso(9)$ on the first three curves $-4,-1,-4$. The Hodge number shifts calculated from F-theory physics (with shared matter representations carefully considered) are $(1, -1)$, which agrees with the Hodge numbers associated with $\Delta$ $(247, 7)+(1, -1)=(248, 6).$

Unlike the previous example, in this case we have tunings 
both on the original fan polytope $(\dd)$
and on the mirror $(\ds)$ side. Nonetheless, the tunings of gauge symmetries in $X$ and
 $\tilde X$ allow us to  use equation
 (\ref{eq:dual-base-tuned}) to
 calculate the bases in both cases:\footnote{We can safely use equation
   (\ref{eq:dual-base-tuned}) without going to the general formula
   (\ref{eq:dual-base-general}) in the absence of Tate tunings of
   gauge symmetries $SU(n), n\geq6$ and $SO(n), n\geq 13.$ }
\begin{equation}
\{w_i^{(\tilde B)}\} = V^B(a_6)
 \text{ and } \{v_i^{(B)}\} = V^{\tilde{B}}(\tilde{a}_6),
\end{equation}
where $a_6$  is associated with lattice points in
$\ds$ of the form $(\cdot,\cdot,-1,-1)$, and
where $\tilde{a}_6$  is associated with lattice points in
$\dd$ of the form $(\cdot,\cdot,-3,-2)$ (cf. equations \ref{45} and \ref{46}.)  All the
examples we have considered so far
are cases in which equation
(\ref{eq:dual-base-tuned}) applies to the calculation of both $B$ and
$\tilde B$.  We now consider a case in which the more general formula
(\ref{eq:dual-base-general}) is required.

\subsection{Standard  stacking $F_{10}$-fibered $\nabla$ vs. non-standard $F_{10}$-fibered $\Delta$}
\label{sec:tuning-3}

\begin{figure}
\centering
\begin{subfigure}{.5\textwidth}
  \centering
  \includegraphics[width=7cm]{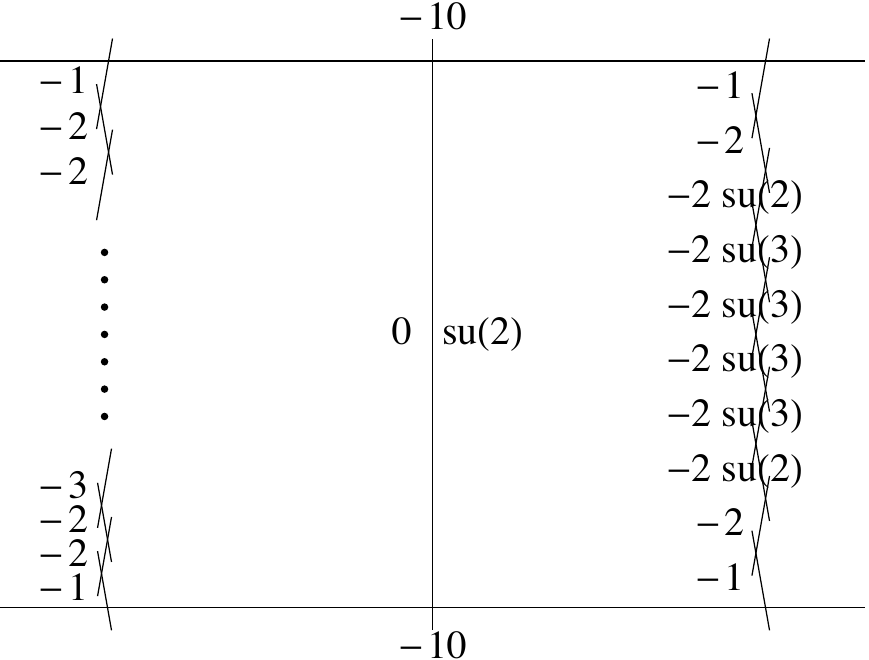}
  \label{fig:b}
\end{subfigure}%
\begin{subfigure}{.5\textwidth}
  \centering
  \includegraphics[width=7cm]{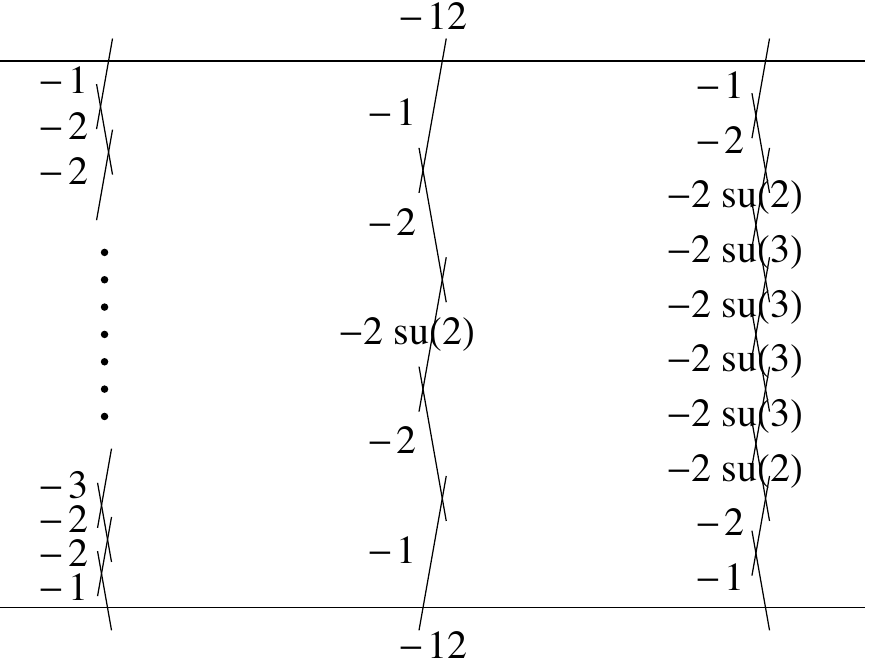}
  \label{fig:a}
\end{subfigure}
\caption{\footnotesize Base resolution that gives rise to a tuned flat elliptically fibered CY. Left: before resolution. Right: after resolving (4,6) points in the base. The top curve is $D_{1}$ and the down curve is $D_{49}$.  The curves in the left chain from top to down are  \{$D_2, D_2$,...,$D_{48}$\}, in the right chain from bottom to top  are \{$D_{50}, D_{51},..., D_{59}$\}.}
\label{fig:nontoric}
\end{figure}

We now consider
tuning an $SU(6)$ on one of the $+ 1$-curves in the $\P^2$ base of the
generic CY elliptic fibration model in \S\ref{sec:example-p2}. The tuned model
corresponds to the polytope M:207 11 N:15 8 H:7,154 in the KS database. The new Hodge numbers
match with the prediction from F-theory physics of tuning the $SU(6)$
on a $+ 1$-curve: $(2,272)+(5,-118)=(7,231).$ The standard stacking
polytope $\nabla$ has vertices
\begin{eqnarray}
\nonumber
\{v_i\}& = &\{ (0,0;1,0),(0,1;-3,-2),(-1,-1;-3,-2),(1,0;-1,-1),(1,0;-3,-2),(0,0;0,1),\\
& & \hspace*{0.1in}(1,0;0,1),(1,0;0,0)\}.
\label{410}
\end{eqnarray} 
The dual polytope $\Delta$ has vertices
\begin{eqnarray}
\{w_i\}&= &\{(-1,-4;-1,0),(-1,-2;-1,1),(-1,-1;0,0),(-1,2;0,0),
(-1,3;-1,1),(-1,5;-1,0),\nonumber\\&&(0,-6;-1,-1),(0,0;-1,2),(0,0;1,-1),(0,6;-1,-1),(12,-6;-1,-1)
\}.
\label{eq:p2-tuning-dual}
\end{eqnarray}
This is however not a standard stacking $\P^{2,3,1}$ polytope,
which can be seen from the feature that there is more than one
monomial in the coefficients of
$x^3$ or $y^2$ in the Tate form
\cite{Huang-Taylor-1}: two lattice points (0,0,0,1) and
(1,0,0,1) in (\ref{410}) contribute to the $x^3$ terms. The set of rays in
the dual base $\tilde B$ is given by equation
(\ref{eq:dual-base-general}), and in this case
\begin{equation}
 \cup_{n  \in\{1, 2, 3, 4, 6\}}V^B  (a_n) = V^B (a_4) \cup V^B (a_6), \,
\end{equation}
which gives a 2D toric fan with the self-intersection numbers
\begin{eqnarray}
\tilde B \rightarrow [[-10//-12//-11//-12//-10,-1,-2,-2,-2,-2,-2,-2,-2,-2,-1]].\nonumber
\end{eqnarray}
The generic elliptic fibration over $\tilde B$ has Hodge numbers (143,
23), so $\tilde X$ is a tuned model.  As $\Delta$ is not a standard
stacking $\P^{2,3,1}$ polytope, we do not have a Weierstrass model
from a Tate form for this polytope. Nonetheless, the Weierstrass model
can be obtained with the trick of ``treating $\Delta$ as a ${\rm
  Bl}_2\P^{1,1,2}$-fibered polytope'' as described in section 4.2 in
\cite{Huang-Taylor-1}.     This is a tuned Weierstrass model with tunings
of gauge symmetries
$\gsu(2)\oplus\gsu(3)\oplus\gsu(3)\oplus\gsu(3)\oplus\gsu(3)\oplus\gsu(2)$
enhanced on $-2, -2,-2,-2,-2,-2$ ($D^{(B)}_{52} - D^{(B)}_{57}$), and
also an $\gsu(2)$ gauge symmetry enhanced on a non-toric 0-curve  intersecting the two $-10$-curves. The polytope $\Delta$ gives a non-flat elliptic fibration model; we can however find an equivalent flat elliptic fibration description with the same tuning of gauge symmetries  over a resolved base  \cite{Huang-Taylor-1}. In the resolved base, the original $-10$-curves are resolved to $-12$-curves through four successive blowups, and the non-toric 0-curve is replaced by curves $-1,-2,-2,-2,-1$ where the
two $-1$-curves intersect with the two $-12$-curves,
respectively (see figure \ref{fig:nontoric}).  The SU(2) gauge symmetry that was enhanced on the non-toric 0-curve is now enhanced on the middle $-2$-curve in the blowup sequence; this is therefore a flat elliptic fibration. The
Hodge number shifts of the flat fibration model calculated from the tunings match exactly with the polytope
model: $(11, -16) + (1,-5) = (154, 7) - (143, 23).$

\subsection{Other toric fibers (Example: vertex stacking on fiber $F_2
  = \P^1 \times \P^1$)}
\label{sec:other-fiber}

For the other examples we have considered so far we have restricted
attention to fibrations with the fiber $F_{10}$ and the ``standard
stacking'' form.  The mirror symmetry structure also factorizes with
other fiber types, though the physics of the corresponding F-theory
models is more complicated and does not follow from standard Tate
tuning structures.  We give one example here of another toric fiber
type, and leave further exploration of mirror symmetry with other
fiber structures to further work.

As a simple example of a mirror pair of elliptically fibered CY3s
associated with fibered polytopes with different fiber types,
we start with an
 $F_2$-fibered polytope with base $B = \P^2$.
We again use the ``stacked'' form where all the rays of the fan of $B$
are embedded within $\dd$ in the form $(v^B_i; v_s), v_s \in\dd_2$.
 Let the
stacking point be one of the $F_2$ vertices $v_s=(1, 0)$. Therefore,
the polytope $\nabla$ has vertices
\begin{equation}
\{v_i\} =\{(-1, -1; 1, 0), (1, 0; 1, 0), (0, 1; 1, 0), (0, 0; 0, 1), (0, 0;   0, -1), (0, 0; -1, 0)\} \,.
\end{equation}
This corresponds to the polytope given by the data M:117 8 N:8 6 H:4,94 [-180] in
the KS database. 

The mirror polytope also has a simple structure.
The dual polytope $\Delta$ has vertices 
\begin{eqnarray}
\nonumber
\{w_i\}& =&\{(-2, -2; 1, 1), (4, -2; 1, 1), (-2, 4; 1, 1),(-2, 4; 1, -1), (-2, -2;   1, -1), (4, -2; 1, -1),\\& &\hspace*{0.1in}
(0, 0; -1, 1), (0, 0; -1, -1)\} \,.
\end{eqnarray}
The mirror fiber $\tilde F  = F_{15}$ has the vertices
\footnote{Note that
$\Delta$ has many distinct fibrations; the numbers of each of the 16 fiber types of $\Delta$ are
$
\{6, 0, 6, 6, 6, 6, 0, 12, 6, 9, 0, 0, 9, 6, 4, 1\}.
$  We can immediately read off the mirror fibration however from the
form of $\dd$, and none of the other fibrations has a corresponding
projection since the fiber of $\dd$ is unique.
}
\begin{equation}
\{(0, 0, -1, -1), (0, 0, 1, -1), (0, 0, 1,
   1), (0, 0, -1, 1)\} \,.
\end{equation}
Over the points in the mirror fiber, we have points associated with
the monomials in ${\cal O}(-2K_B)$ over all the points $(1, y), y =
-1, 0,
1$, and points associated with ${\cal O} (-K_B)$ over the points $(0,
y), y = -1, 0, 1$, and only the points $(0, 0; -1, y)$ over the
remaining points in the fiber.  From this we see that the dual base
 $\tilde B$
is given by the toric surface with the self-intersection sequence
that  can be read off from $\mathcal{O}(-2K_B)$,
\begin{eqnarray}
\nonumber
\tilde B\rightarrow[[-1,-4,-1,-3,-1,-4,-1,-4,-1,-3,-1,-4,-1,-4,-1,-3,-1,-4]].
\nonumber
\end{eqnarray}

While the factorization of mirror symmetry is equally clear in this
example to the others considered here, the F-theory interpretation is
more subtle.  A full analysis involves considerations using methods
like those of \cite{Bouchard-Skarke, Klevers-16}.  We outline the
analysis on the $\dd$ side in this case and leave further work in this
direction to the future.  The polytope $\nabla$ has the single obvious
fiber $F_2$, which does not provide a section, so this is a genus one
fibration.  We can analyze the gauge group and matter structure of the
corresponding Jacobian fibration, which is relevant for F-theory
\cite{Morrison-WT-sections}. 
The Weierstrass model of the Jacobian
fibration has no nonabelian gauge symmetries.  From the Hodge numbers
we expect a nontrivial Mordell-Weil group of rank 2,
\begin{equation}
G=U(1)\times U(1)\,.
\end{equation} 
Codimension two singularities in the Weierstrass model suggest  $36 +
72 + 72$  matter fields charged in various ways under the U(1) factors in
$G$, which is in accordance with the expected Hodge numbers $(4,
94)=(2, 272) + (2, 2 - (36 + 72 + 72)))$.\footnote{Note that
from \cite{Klevers-16},
the F-theory models associated with the Jacobian
  fibrations of $F_2$ fibered polytopes should
generically have
  toric Mordell-Weil group $U(1) \times \Z_2$. 
The stacked form we have here over a vertex of $F_2$ gives a degeneration that may enhance
the $\Z_2$ to a U(1) factor. (Thanks to Paul Oehlmann for explaining
this to us.)
} 

\subsection{Elliptic fibration with a non-fibered mirror}
\label{sec:non-fibered}

As a final Calabi-Yau threefold example we consider a case where a CY
threefold has an elliptic fibration associated with a 2D reflexive
subpolytope, but the mirror has no fibration.  In such cases, there
cannot be a projection onto the fiber of $\ds$, as discussed in
\S\ref{sec:2.2}.

We consider the polytope from the KS database associated with the
Calabi-Yau threefold with Hodge numbers $(h^{1, 1},h^{2, 1}) =(149,
1)$.
There is a coordinate system in which this polytope $\dd$ has vertices 
\begin{equation}
\{v_i\} =\{(0, 0; 1, 0), (-2, 8; -3, -2), (-2, 0; -3, -2), (6, 0; -3,
-2), (-2, -8, 5, 6)\} \,.
\end{equation}
In this coordinate system there is an $F_{10}$ reflexive 2D fiber in
the standard form $(0, 0; v^F)$.  The polytope does not
satisfy the condition needed for the mirror to have a fibration,
however, since the last vertex $v= (v^{(I)}; v^{(II)}), v^{(II)} = (5,
6)$ does not satisfy $v^{(II)} \in\dd_2$.  Furthermore, it is
straightforward to see that no linear transformation that preserves
$\dd_2$ can move all the vertices to satisfy this condition.  In
particular, the second and last vertices will always have $v^{(II)}$
values that differ by a vector of the form
\begin{equation}
v^{(II)}_5-v^{(II)}_2 = (8, 8) + 16 (x, y), x, y \in\Z \,.
\end{equation}
Thus, the mirror can never satisfy the fibration condition.  This is
not surprising since the mirror has Hodge numbers (1, 149) and cannot
have a genus one or elliptic fibration since the Shioda-Tate-Wazir
would give $h^{1, 1}(\tilde{X}) \geq h^{1, 1} (\tilde{B}) + 1 \geq 2$.
One can also check explicitly that the lattice points in the mirror
polytope $\ds$ do not contain a linearly embedded $\P^{2, 3, 1}$ fiber.

\subsection{Elliptic Calabi-Yau fourfolds (Example: generic elliptic
  fibration over $\P^3$)}
\label{sec:4D}

The factorization structure that we have described for elliptic toric
hypersurface Calabi-Yau threefolds can occur in much the same fashion
for higher-dimensional Calabi-Yau manifolds realized as toric
hypersurfaces.  We leave a more detailed investigation of
higher-dimensional Calabi-Yau mirror symmetry for future work, but
note here simply that the simplest cases of generic elliptic
fibrations over toric bases have a natural generalization to arbitrary
higher dimension.  We present here only a single simple case, the
 Calabi-Yau fourfold given by the generic elliptic fibration over the
 base $\P^3$.  The polytope $\nabla$ in this case is the simple
 generalization of the polytope with vertices
given by Eq.\
(\ref{eq:p2-nabla}), and has vertices
\begin{align}
\{v_i\}  & =\{(0, 0, 0; 1, 0), (0, 0, 0; 0, 1), (1, 0, 0; -3, -2), (0, 1,
0; -3, -2),\\&\hspace*{0.1in} (0, 0,
1; -3, -2),
(-1, -1, -1; -3, -2)\} \,.
\label{eq:p3-nabla}
\end{align}
The Hodge numbers of the corresponding Calabi-Yau fourfold are $h^{1,
  1}(X)= 2,h^{3,1}(X) = 3878$.
The dual polytope $\Delta$ is again $\P^{2, 3, 1}$ fibered, with a base
$\tilde{B}$ given by the set of toric rays within the polytope having vertices
\begin{equation}
\{v^{(\tilde{B})}_i\} =\{(-6, -6, -6), (18, -6, -6), (-6, 18, -6), (-6, -6,
18)\} \,.
\end{equation}
There are a tremendous number of ways of triangulating this
5D polytope to
give a complete toric fan.  Independent of the triangulation, however,
the generic elliptic fibration over the toric base $\tilde{B}$ will
have a non-Higgsable gauge group
\begin{equation}
G = E_8^{34} \times F_4^{96} \times G_2^{256} \times SU(2)^{384}\,.
\end{equation}
This can be seen from the numbers of primitive
rays in $\tilde{B}$ with specific minimum values of the inner products with the
vertices in (\ref{eq:p3-nabla}).  
For each triangulation, the mirror Calabi-Yau will have Hodge numbers
$h^{1, 1}(\tilde{X}) = 3878,h^{3,1} (\tilde{X}) = 2$.
An interesting feature of the mirror
Calabi-Yau fourfolds associated with the polytope $\Delta$ is that it
is one of the ``attractive'' endpoints reached by blowing up $\P^3$ as
far as possible consistent with a base that supports an elliptic
fibration \cite{Wang-WT-MC-2}. This may be associated with the large
number of triangulations that are possible in this case.  The other
attractive endpoints also are factorizable mirrors over simple bases
that are 3D toric varieties with few rays.  We leave further
exploration of the many interesting questions associated with mirror
symmetry and Calabi-Yau fourfolds to further work.

\section{Conclusions and further questions}
\label{sec:conclusions}

\subsection{Summary of results}

Building on recent work \cite{Huang-Taylor-2} in which we showed that
most toric hypersurface Calabi-Yau threefolds have a manifest genus
one or elliptic fibration, we have found that many of these elliptic
fibrations exhibit a mirror symmetry that factorizes, in the sense
that the fiber $F$ of $X$ is a mirror Calabi-Yau 1-fold to the fiber
$\tilde{F}$ of $\tilde{X}$.  This connects with a number of directions
of earlier research related to aspects of such mirror fibers
\cite{Avram:1996pj, Berglund-Mayr, Grassi-P, Klevers-16, Braun-gk,
  Oehlmann-rs, Cvetic-gp}.
We have furthermore found that the structure of the mirror base
$\tilde{B}$ is determined in a clear and well-defined way from the
base $B$ and fibration structure of $X$.  In many cases of interest
the mirror base takes a simple form in terms of the toric geometry of
a line bundle over $B$.  In particular, for generic
CY elliptic
fibrations over toric base surfaces, the mirror base $\tilde{B}$ has a
toric fan that is built from the primitive rays in the set of sections
of the line bundle ${\cal O} (- 6 K_B)$.
For tuned Tate models over a toric base surface, there is a slightly
more complicated expression for the rays in $\tilde{B}$, given by
(\ref{eq:dual-base-tuned}) when $\Delta$ is also a standard stacking polytope, or (\ref{eq:dual-base-general}) when $\Delta$ fails to maintain the standard stacking structure due to excessive removal of points associated with monomials in $\mathcal{O}(-6K_B)$. As shown
in \cite{Huang-Taylor-1}, almost all the Hodge number pairs in the KS
database with $h^{1, 1}\geq 240$ or $h^{2, 1}\geq 240$ are realized by
generic or tuned elliptic fibrations with the standard stacking structure.

We have explored some simple examples of this factorized mirror
symmetry, particularly for some cases of Calabi-Yau threefolds with
large Hodge numbers, and generic and tuned
elliptic fibrations over simple bases
such as $\P^2$ and the Hirzebruch surfaces $\F_n$.

With growing evidence that most known Calabi-Yau threefolds admit a
genus one or elliptic fibration, the results we have found here
suggest that there may be a very general way of understanding mirror
symmetry in terms of fibration by smaller-dimensional Calabi-Yau
fibers.
Further work is clearly needed to explore the details of the mirror
dictionary for different bases and fibers, in higher dimensions, and
the extent to which the factorization structure identified here can be
extended beyond the toric hypersurface framework.

\subsection{Further questions and directions}

We list here some specific open questions that may be of interest for further research.

$\bullet$ We have given a variety of examples here where mirror
symmetry between a pair of elliptic Calabi-Yau threefolds factorizes
between the base and the fiber of the fibrations.  From these examples
it is clear that there are regular local structures that could be used
to begin to form a dictionary relating structure on the mirror base
$\tilde{B}$ to structure of the base $B$ and fibration structure of
the original elliptic Calabi-Yau $X$.  For example, a curve of
self-intersection $+1$, such as was encountered in the example in
\S\ref{sec:example-p2}, naturally corresponds to a sequence of toric
curves of self-intersections $//-12//-12//$ in the mirror base
$\tilde{B}$. Similarly, in Table~\ref{f:Hirzebruch-bases-nontoric}, we
see a pattern where blowing up a non-toric vs.\ toric point in the base $B$
corresponds to blowing up a toric vs.\ non-toric point in the mirror base
$\tilde{B}$.  It would be interesting to try to systematically develop
this kind of structure, ideally including the additional reductions on
the base $\tilde{B}$ that are imposed by different Tate tunings on the
generic elliptic fibration over a toric $B$, which can be understood
through additional constraints arising from the associated ``tops''.

$\bullet$ We know that at large Hodge numbers many of the Calabi-Yau
threefolds in the Kreuzer-Skarke database are generic or tuned
elliptic fibrations over toric bases that can be constructed from
polytopes with a fiber $F_{10}$, and these exhibit the simplest forms
of mirror factorization studied in \S\ref{sec:generic},
\S\ref{sec:further-examples}.  It would be interesting to study
further what fraction of the KS database exhibit mirror factorization
and what other types of structures arise frequently or in isolated
cases at smaller Hodge numbers.

$\bullet$ We have focused here primarily on the simplest cases where
the toric 2D fiber is the self-dual fiber $F_{10}$, corresponding to
generic elliptic fibrations.  It would be interesting to study in more
detail the structure of the other 2D toric fibers.  In particular, it
was found in \cite{Klevers-16} that the dual fibers $F_i,
\tilde{F}_i$ exhibit some interesting structure, including identical
numbers of sections associated with toric Mordell-Weil rank, and a
matching between Mordell-Weil torsion on one side and discrete
symmetries associated with the Tate-Shafarevich/Weil-Chatalet group on
the dual side.  It would be interesting to understand better how these
features of the fibers can be used in understanding mirror symmetry of
the full Calabi-Yau threefolds $X, \tilde{X}$ with the different dual
fiber types.

$\bullet$ One natural way of trying to extend the analysis here is to
look at complete intersection Calabi-Yau varieties.  A large class of
complete intersection fibers were analyzed in \cite{Braun-gk}, and the
properties of mirror fibers in these more general cases were noted in
this paper and studied more thoroughly in \cite{Oehlmann-rs}.  It
would be interesting to investigate these structures further in the
context of full elliptic Calabi-Yau threefolds (and fourfolds).

$\bullet$ One of the most powerful approaches to mirror symmetry that
has been used in earlier work is the Strominger-Yau-Zaslow (SYZ)
picture \cite{SYZ}, in which mirror symmetry is realized by T-duality
on a 3-torus fiber over a real threefold base.  While this picture has
led to some powerful insights into mirror symmetry, it is incompatible
with the algebro-geometric structure of Calabi-Yau manifolds, and the
3-torus fibration of a general Calabi-Yau is extremely singular.  The
factorization structure identified here seems to match more naturally
with ideas from algebraic geometry and involve more controlled
singularity structures.  It would be interesting to understand whether
there is a way of relating the SYZ picture to the factorization
structure found here. 

$\bullet$ As described in a simple example in \S\ref{sec:4D}, the
factorization structure explored here should be equally valid for
higher-dimensional Calabi-Yau varieties, and particularly for
Calabi-Yau fourfolds.  It would be interesting to explore further the
structures that arise for mirror symmetry of elliptic Calabi-Yau
fourfolds.

$\bullet$ The mirror symmetry identified here between elliptic
fibrations suggests that there may be an interesting corresponding
duality in F-theory.  This would be interesting to explore further.

$\bullet$  It would be interesting to connect this factorization
structure of mirror symmetry to other aspects of mirror symmetry
research and Calabi-Yau geometry.  For example, it would be
interesting to understand how the form of Calabi-Yau periods and the
structure of the moduli space, or recent progress  on the all genus amplitudes of topological string theory on elliptic Calabi-Yau threefolds \cite{hkk} fits into this factorized picture.

\acknowledgments{Particular thanks to Yinan Wang for insight and discussions related to
this work; as mentioned in \S\ref{sec:generic}, the simplest class of
examples of the mirror symmetry structure explored here were
identified some years
earlier by Wang in the context of a separate project with Andreas Braun
\cite{btw-unpublished}.
We would also like to thank 
Lara Anderson, Andreas Braun, Antonella Grassi, James Gray, Sheldon Katz, Albrecht Klemm, Dave Morrison, Paul
Oehlmann,
and Andrew Turner
  for helpful discussions. 
This material is based upon work supported by the U.S.\ Department of
Energy, Office of Science, Office of High Energy Physics under
grant Contract Number
DE-SC00012567.
}

\newpage
\appendix

\section{The 16 reflexive 2D fiber polytopes $\dd_2$}
\label{sec:appendix-fibers}

We list here the 16 reflexive 2D polytopes $\dd_2$.  
The mirror of fiber $F =F_i$ is the fiber $\tilde{F} = F_{17-i}$ for
$i < 7, i > 10$; the fibers $F_i$ for $i = 7, 8, 9, 10$ are
self-mirror up to linear transformations.
For each lattice point $v^F$, we
include the maximum value of $1 +v^F \cdot m^{(II)}, m^{(II)} \in\ds_2$.

\vspace*{0.2in}
\begin{center}

{\centering
\begin{tabular}{cccc}
\includegraphics[height=3.3cm]{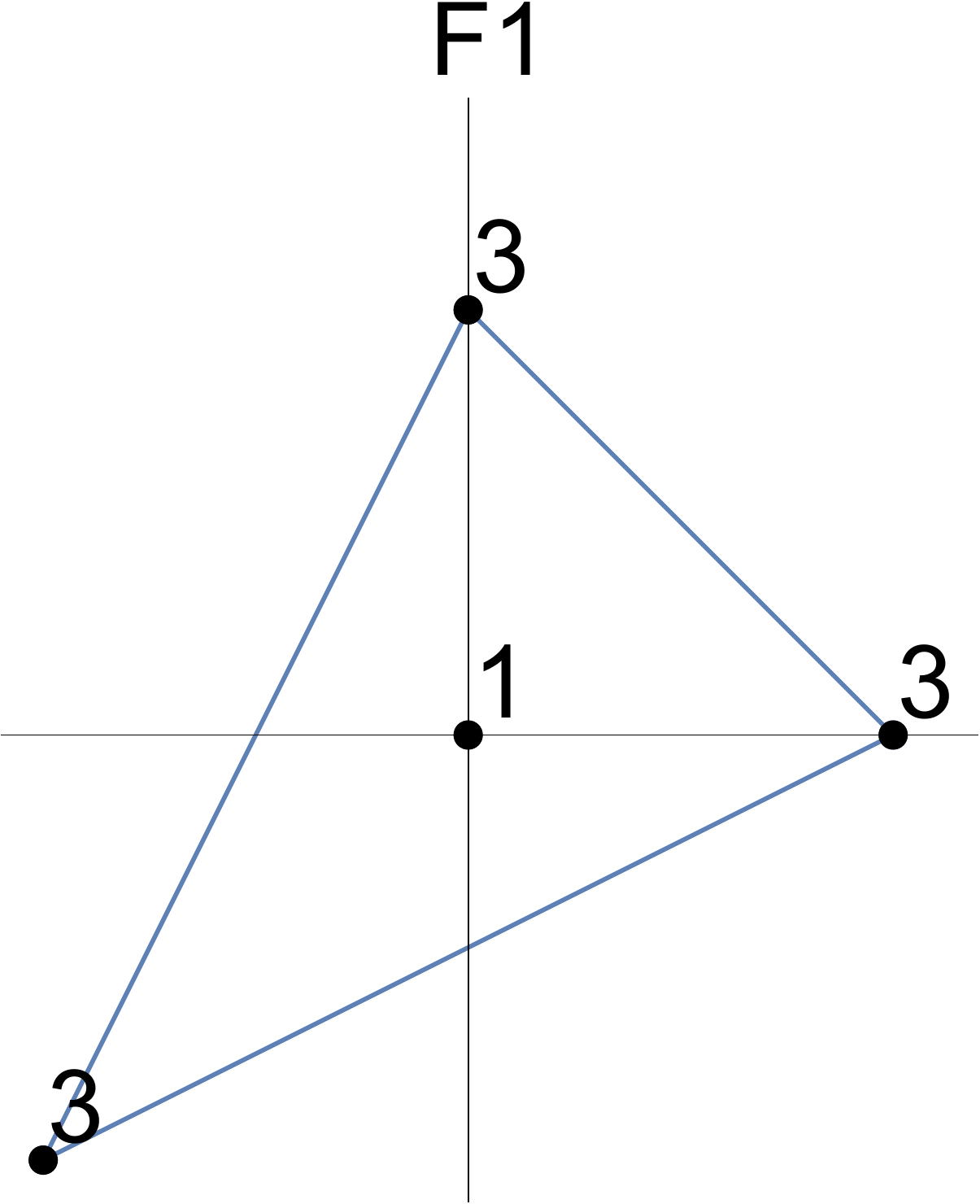}  &\includegraphics[height=3.3cm]{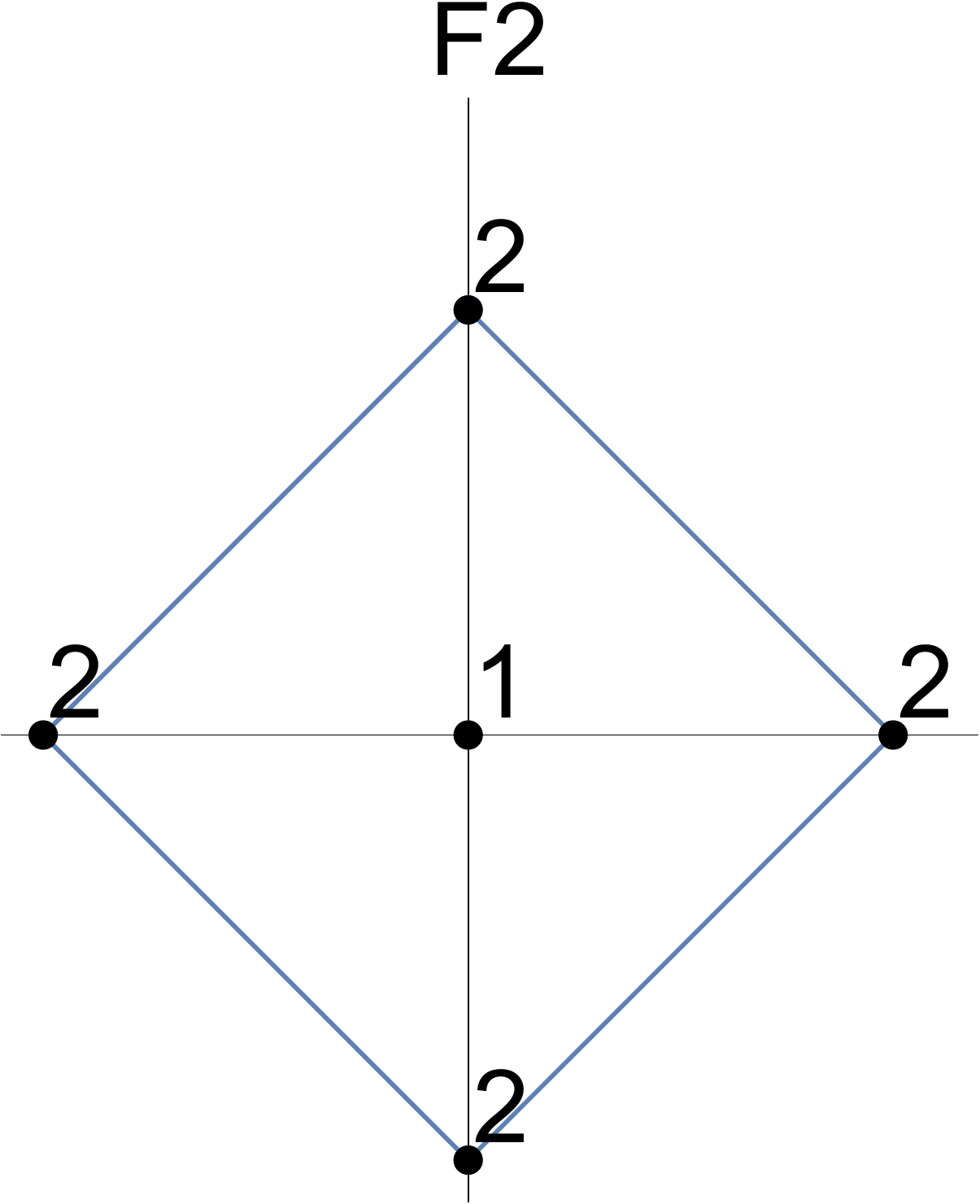}    & \includegraphics[height=3.3cm]{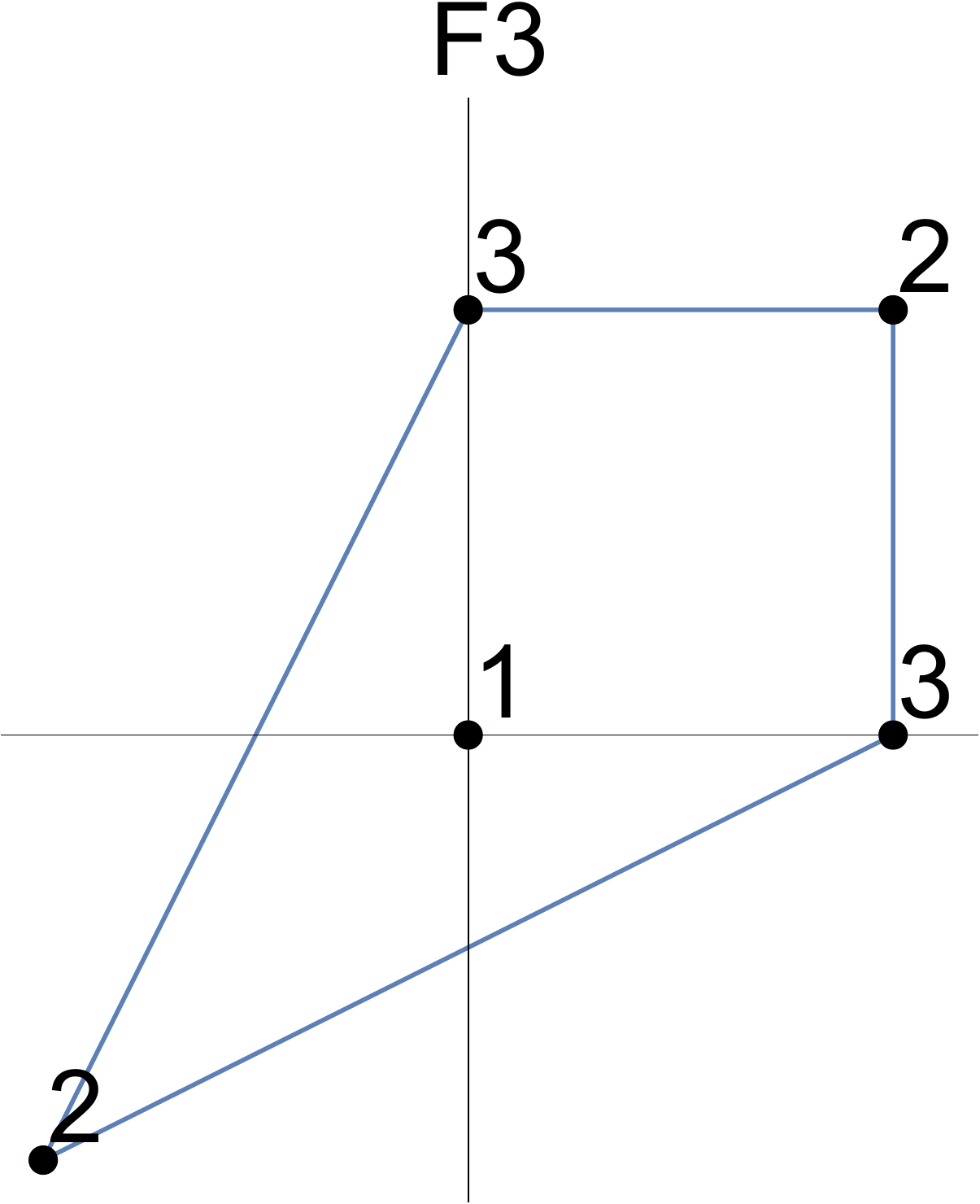}   & \includegraphics[height=3.3cm]{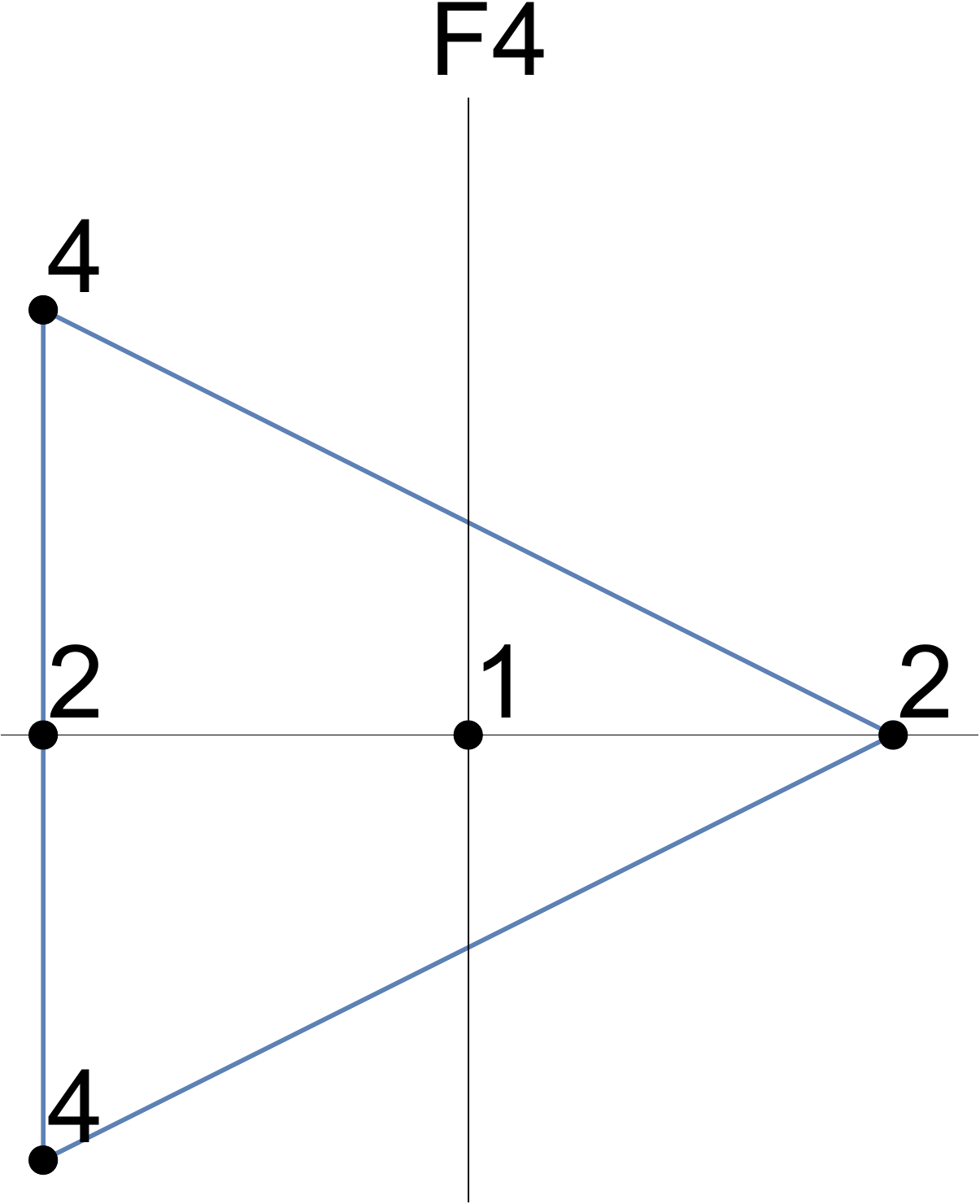}   \\\\
\includegraphics[height=3.3cm]{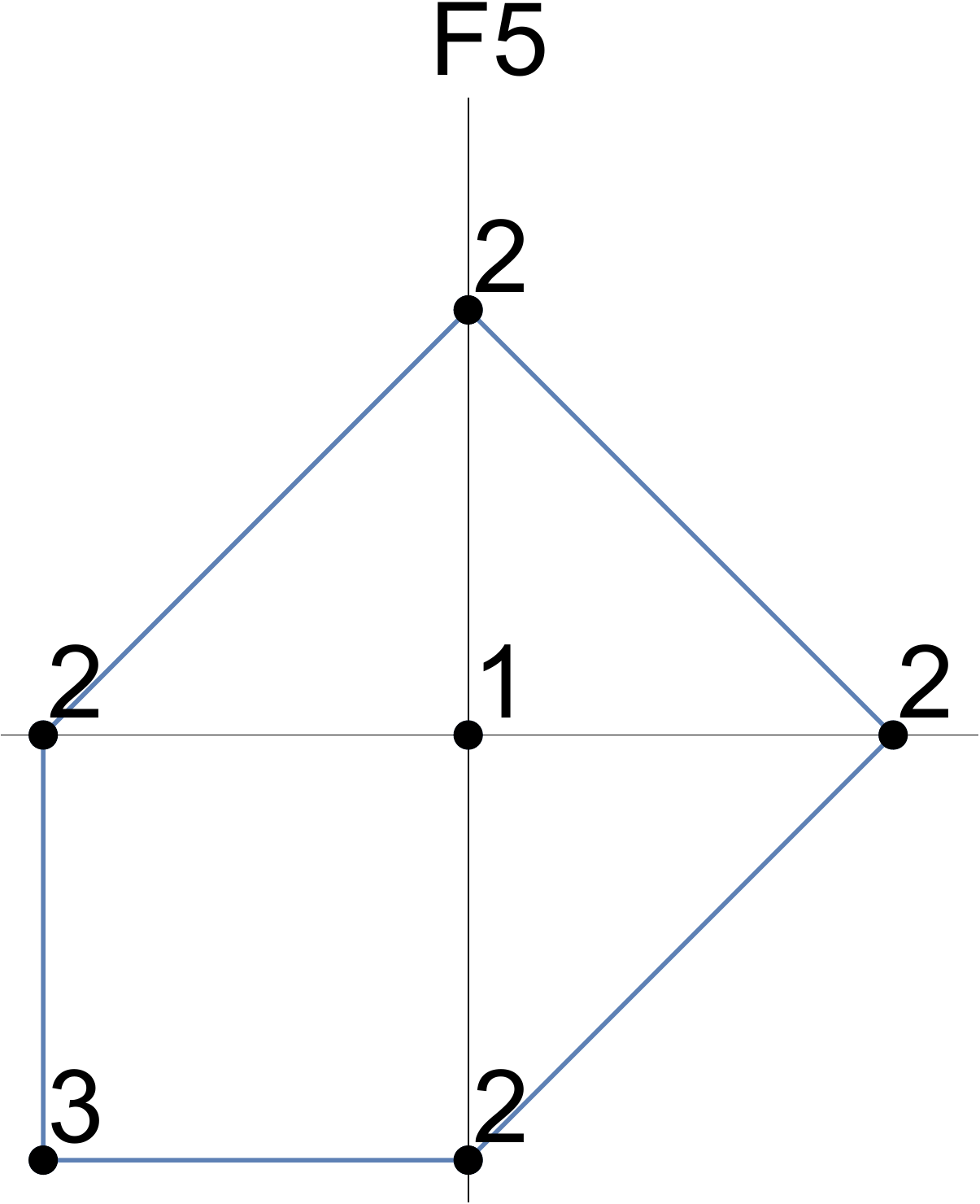}          & \includegraphics[height=3.3cm]{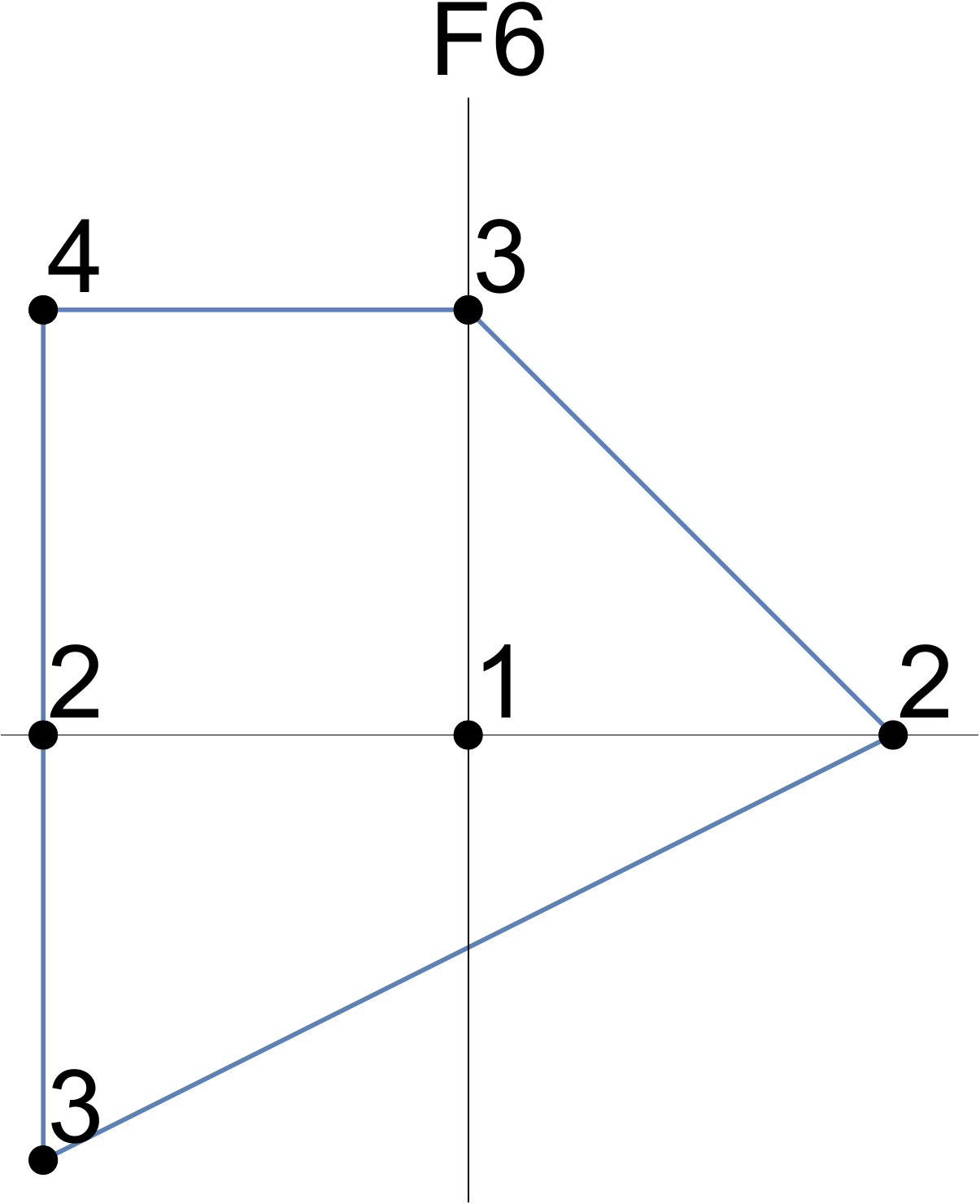}       &\includegraphics[height=3.3cm]{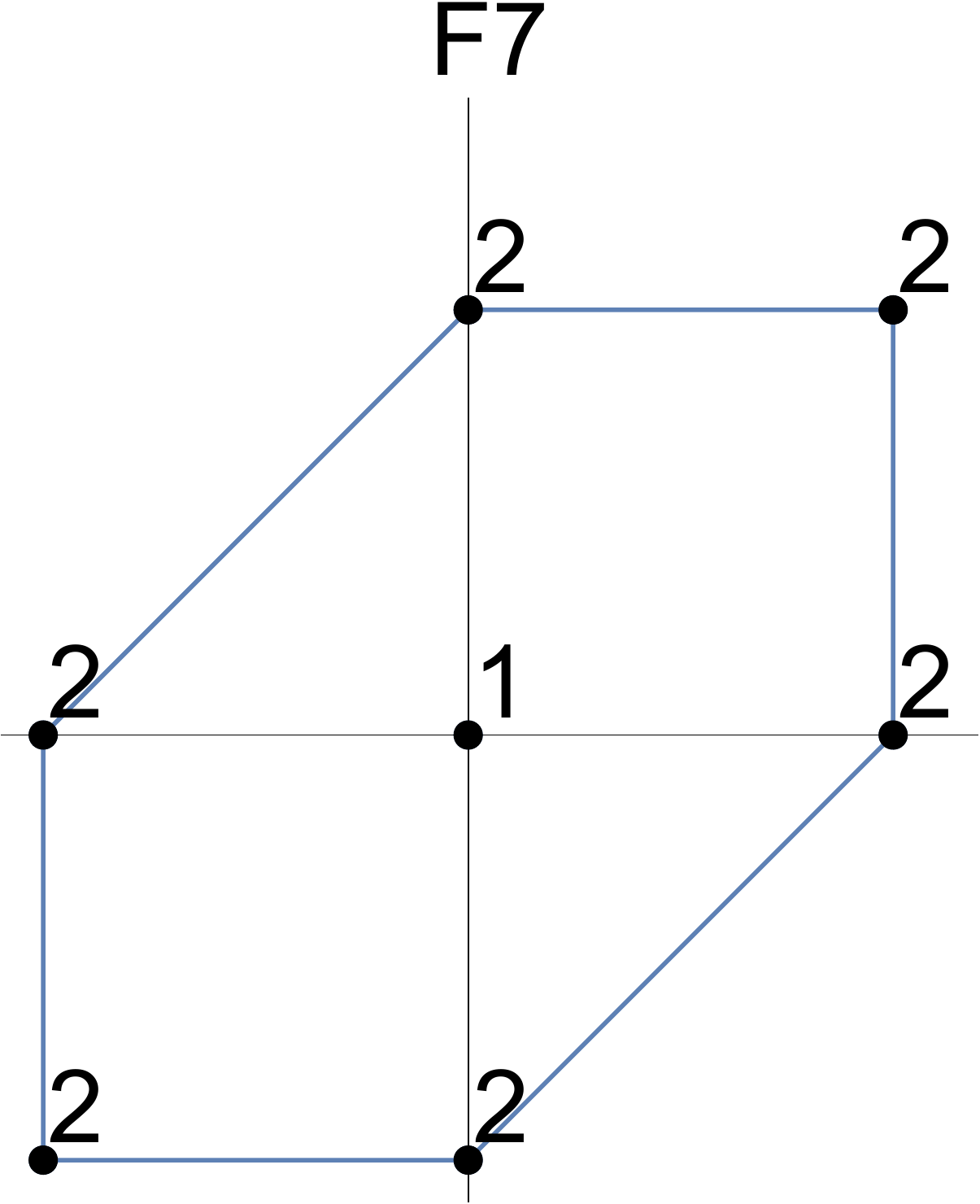}        &\includegraphics[height=3.3cm]{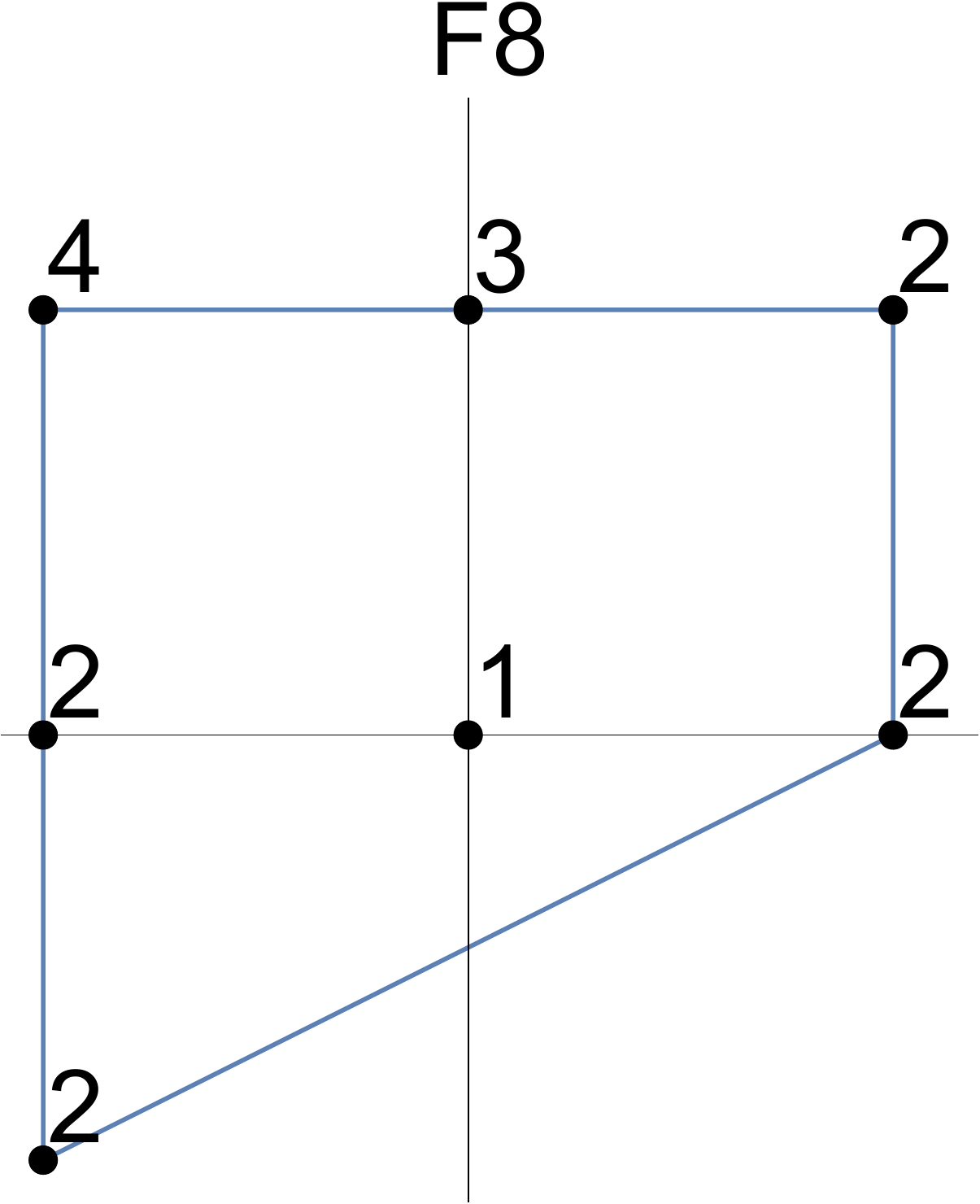}        \\\\
\includegraphics[height=3.3cm]{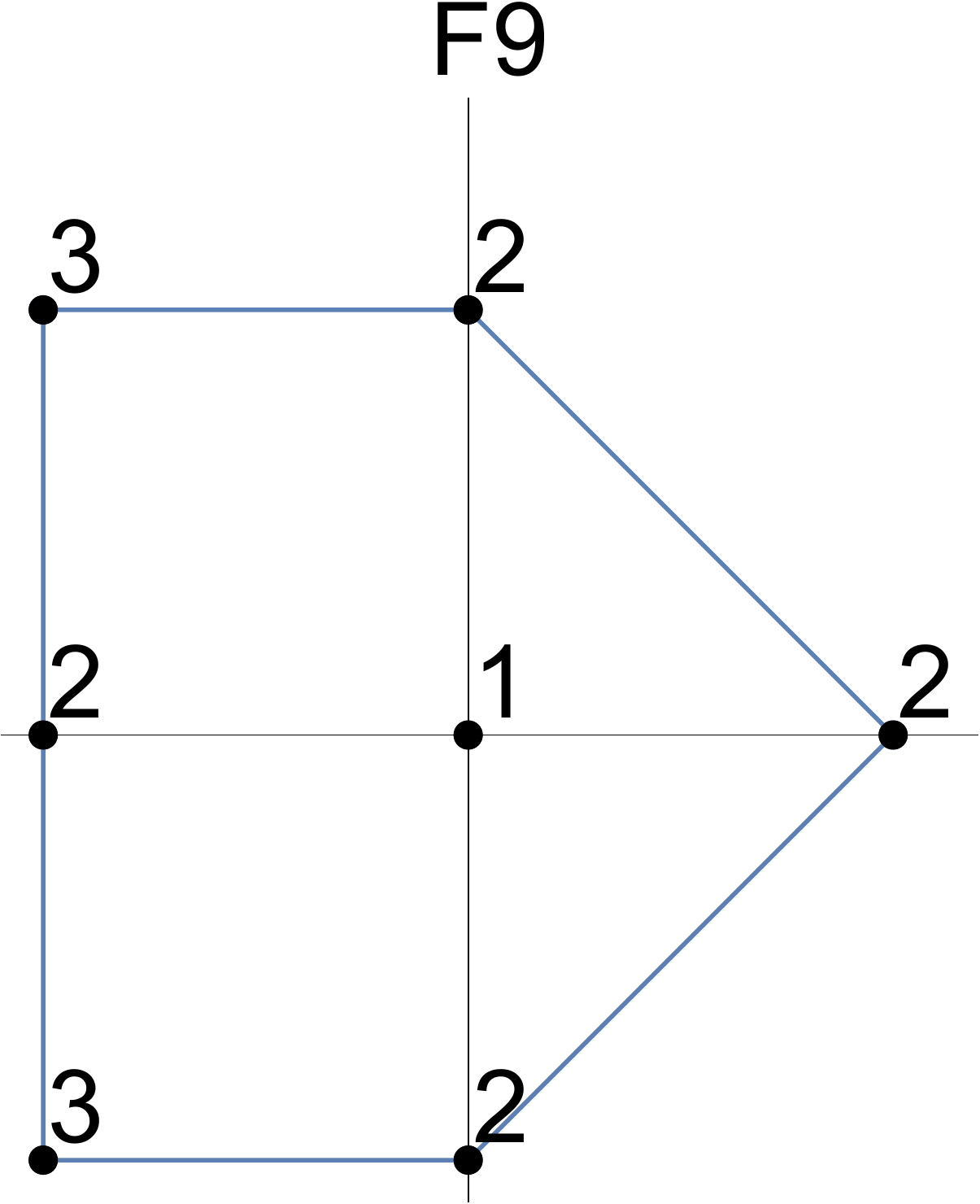}          & \includegraphics[height=3.3cm]{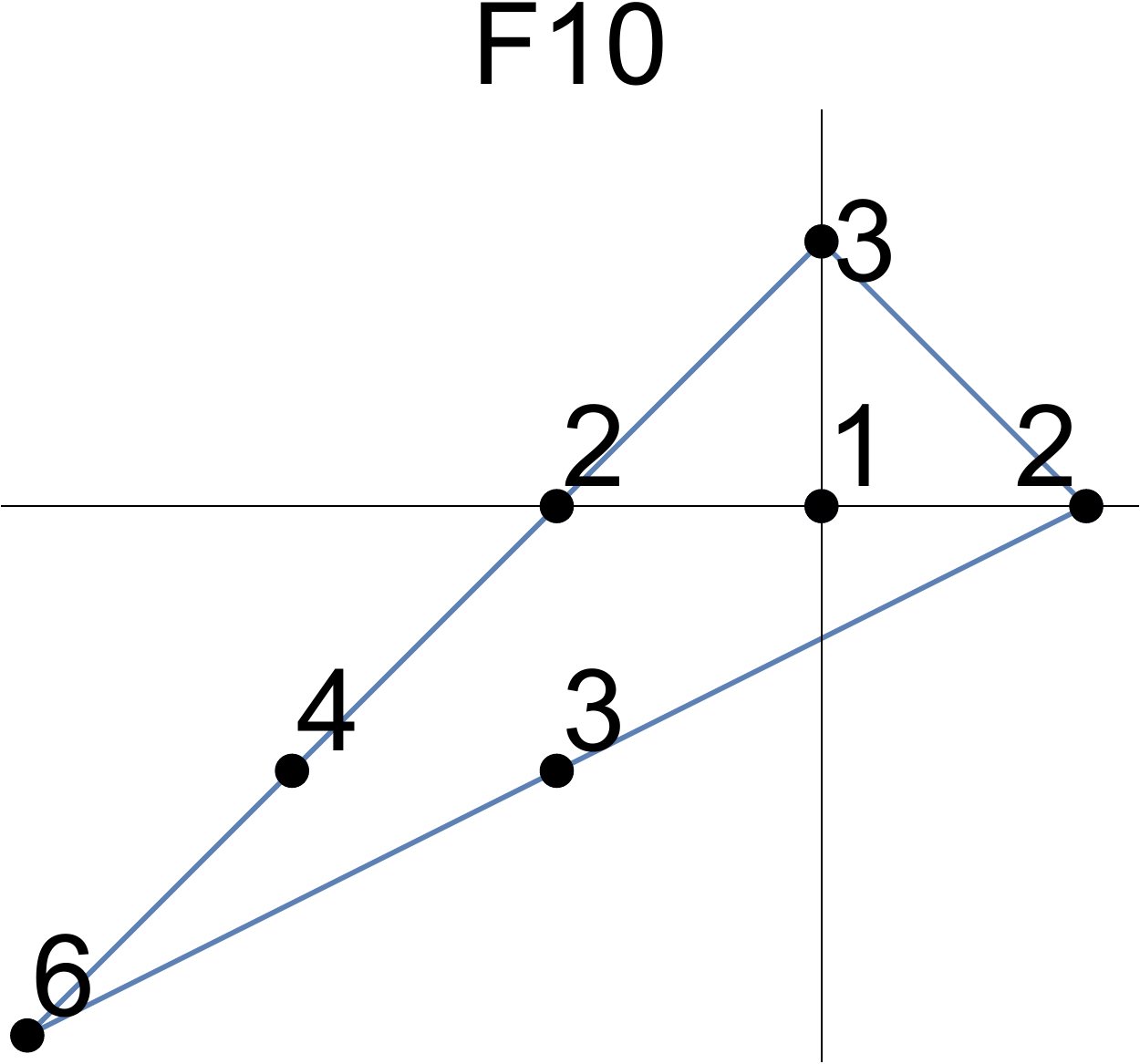}       &\includegraphics[height=3.3cm]{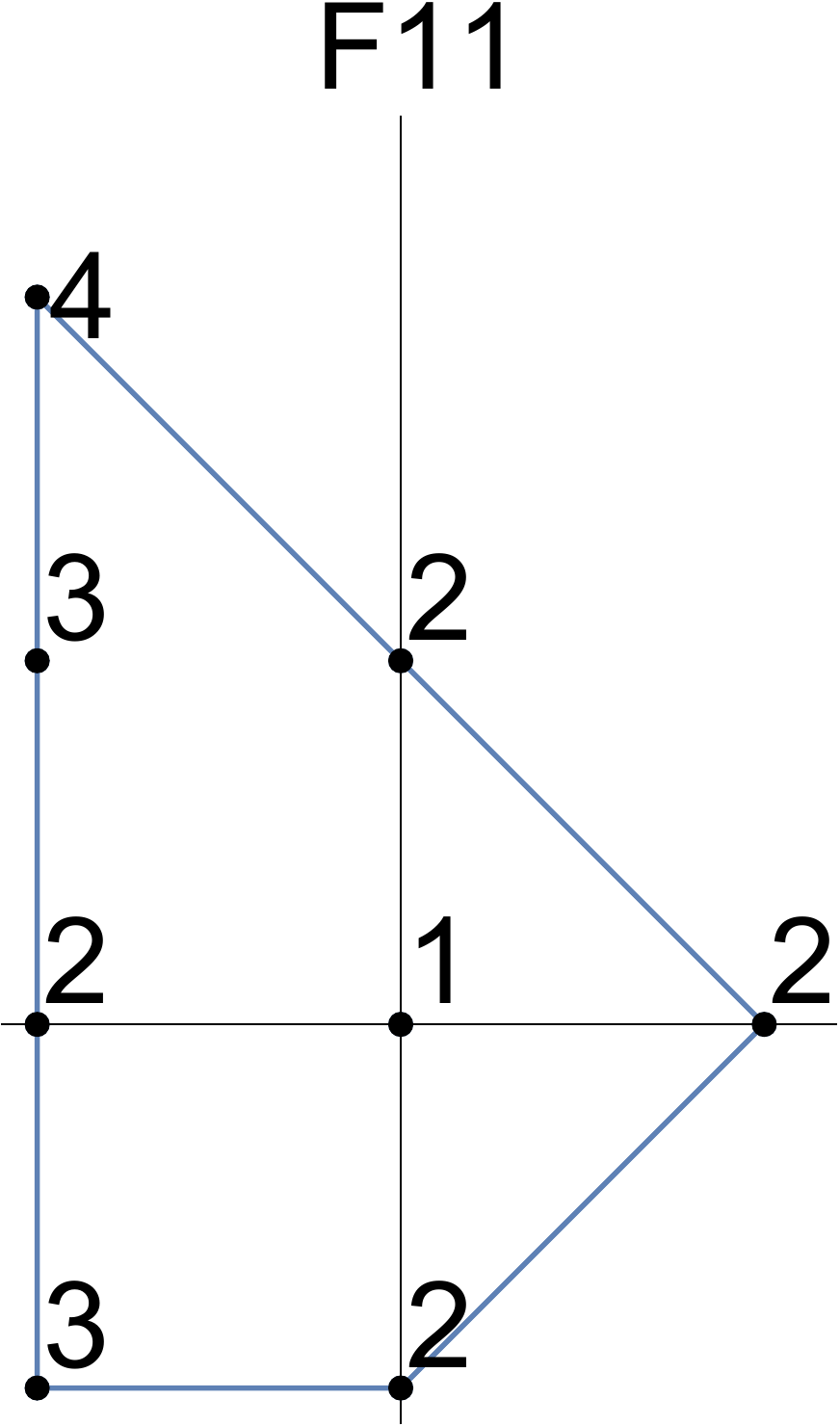}        &\includegraphics[height=3.3cm]{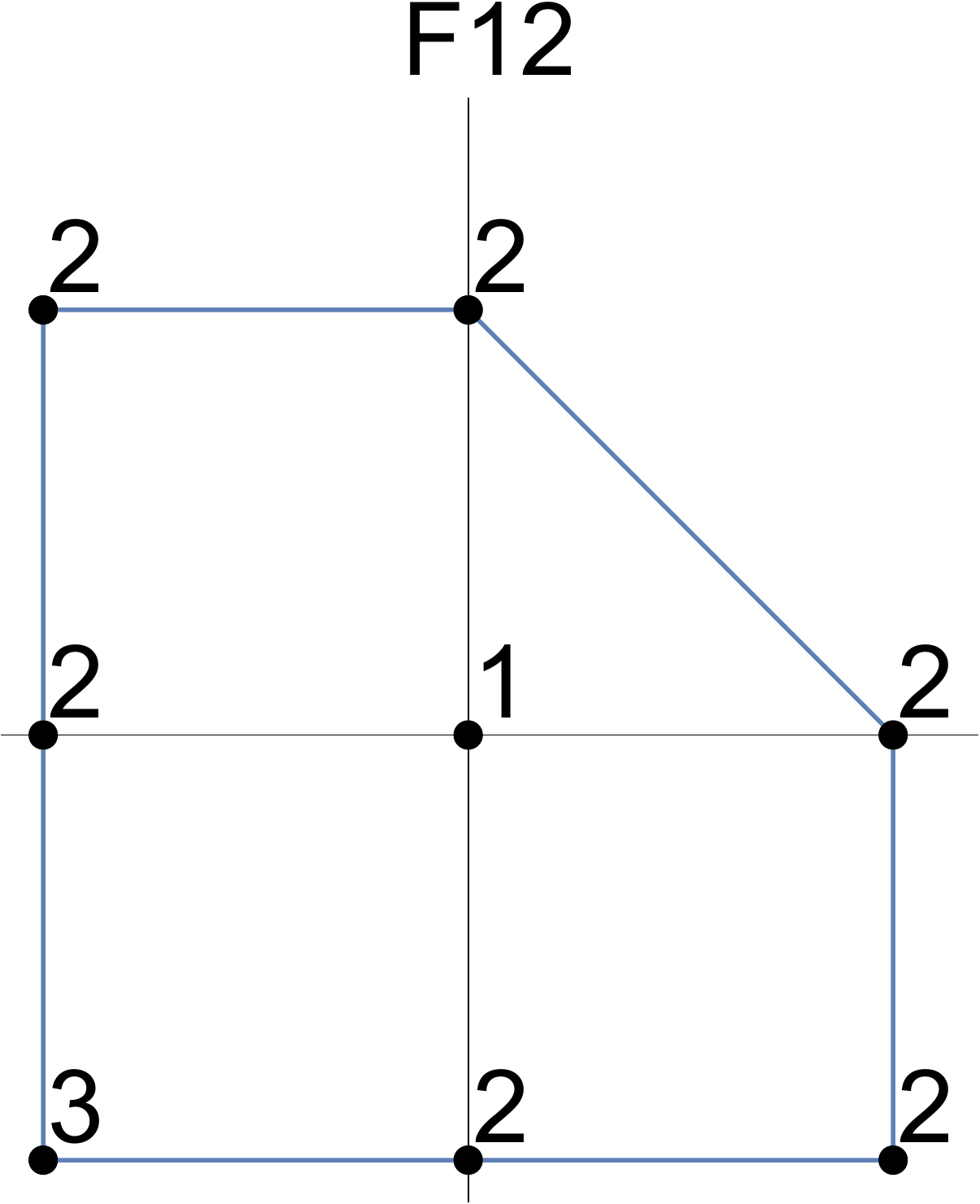}        \\\\
\includegraphics[height=3.3cm]{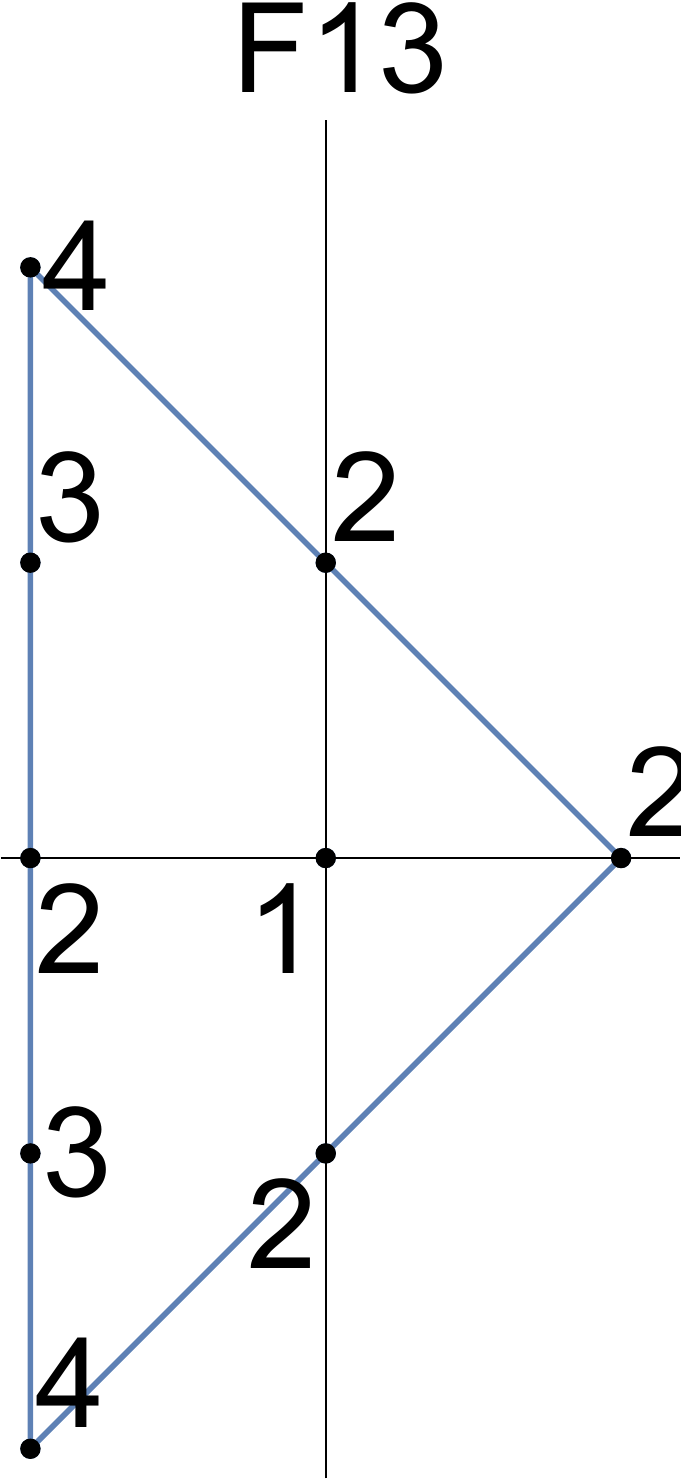}          & \includegraphics[height=3.3cm]{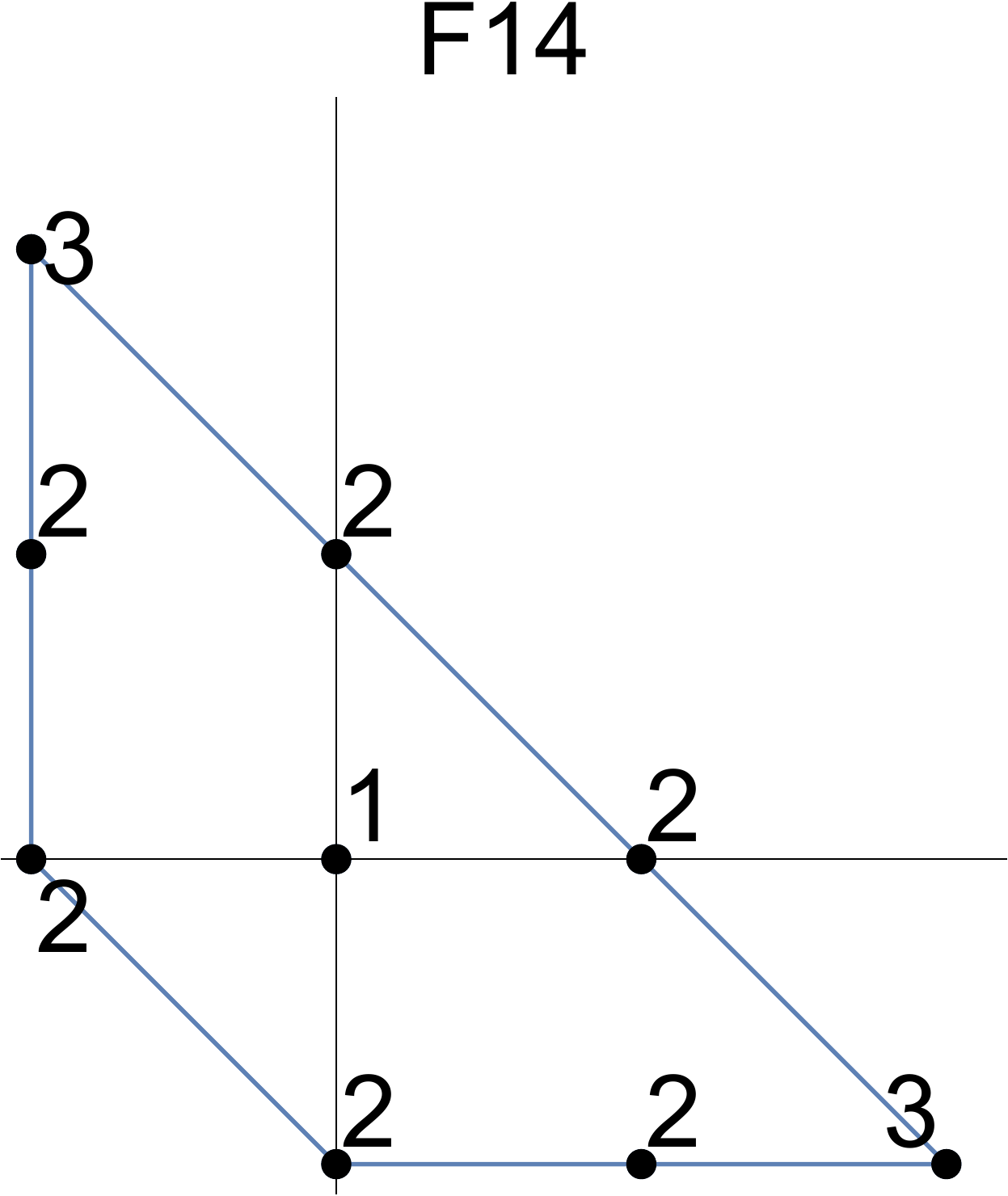}       & \includegraphics[height=3.3cm]{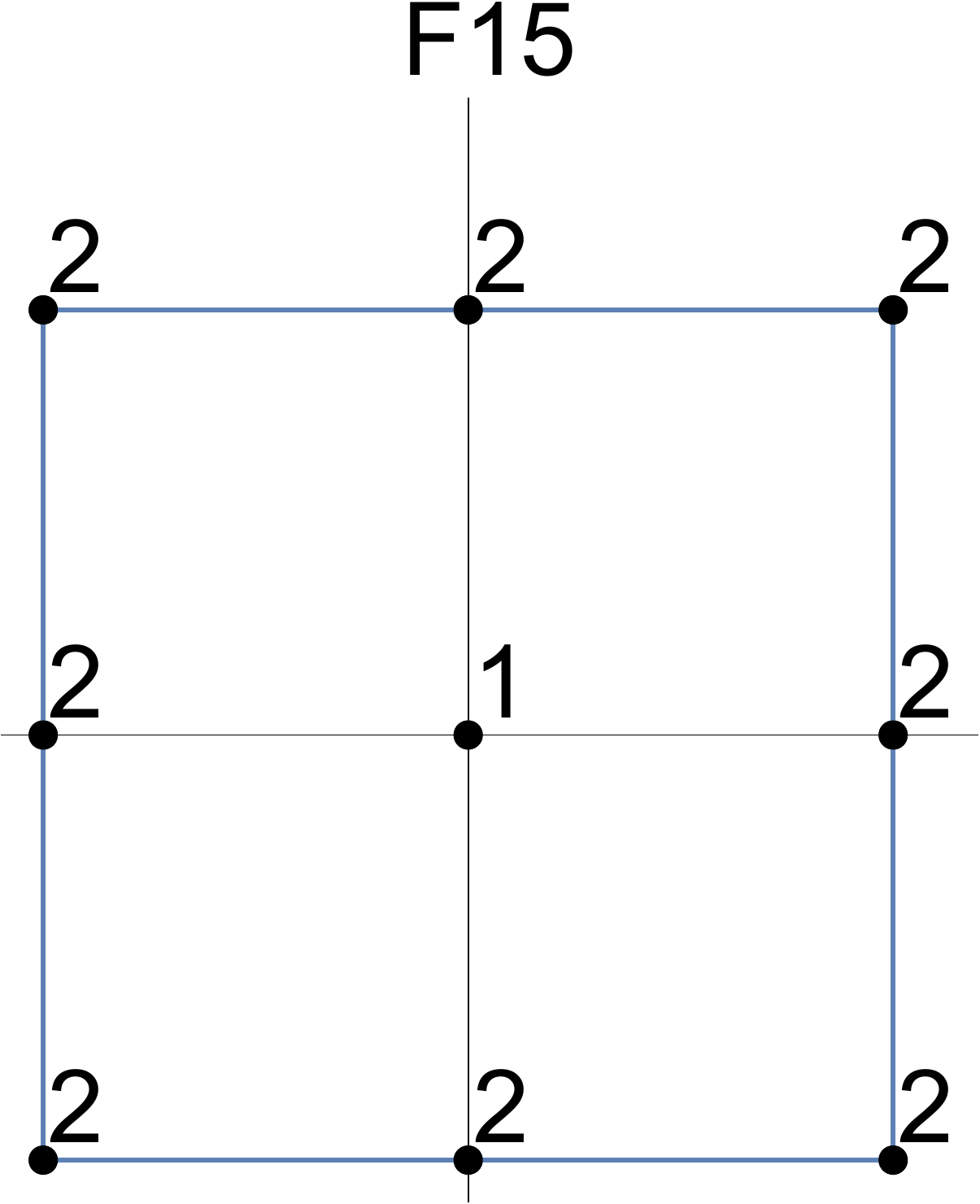}       &\includegraphics[height=3.3cm]{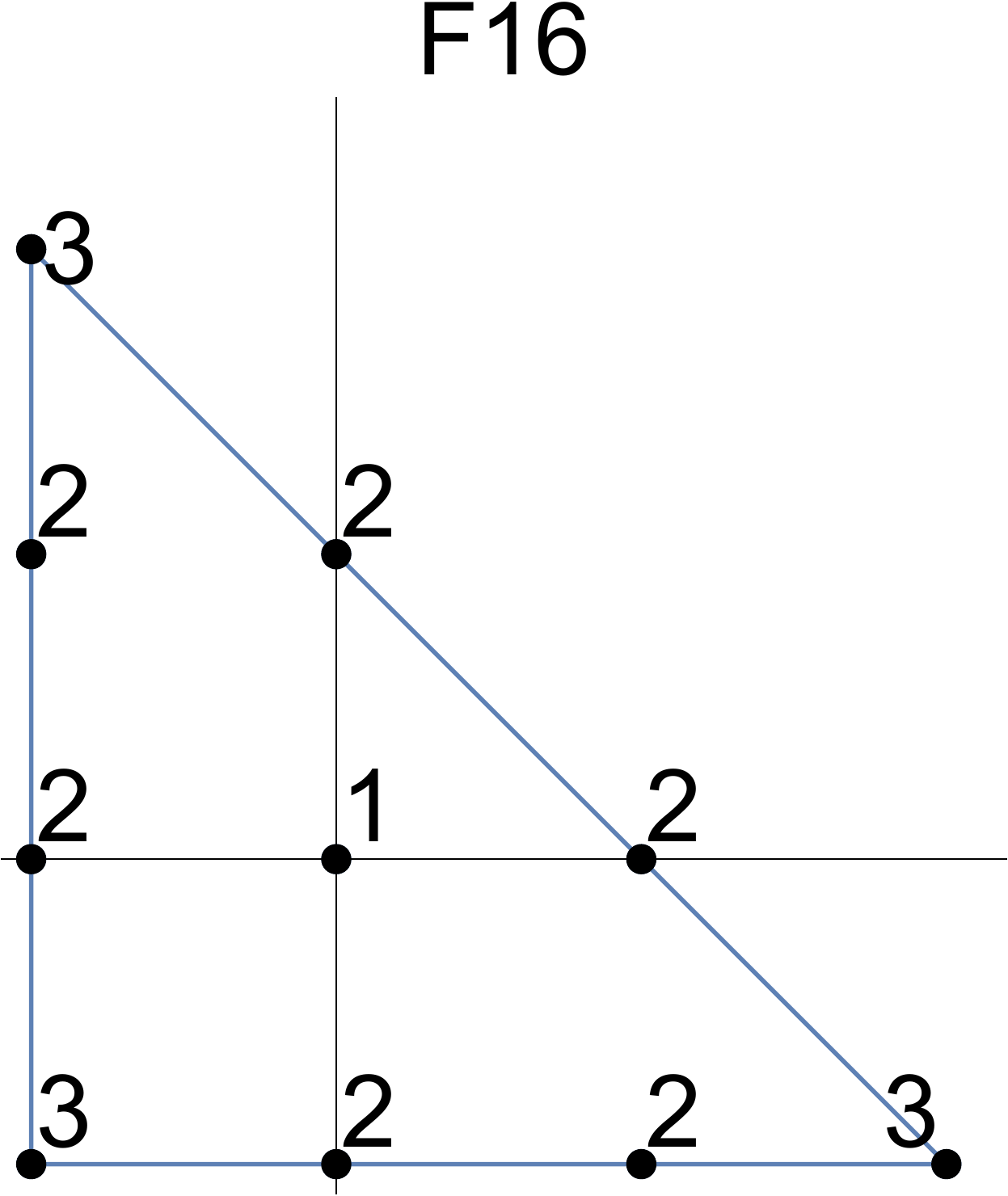}       
\end{tabular}
}
\end{center}
\section{Faces of the base polytope and  chains of
  non-Higgsable clusters}
\label{sec:chains}

Certain chains of self-intersections of curves in the base, associated
with characteristic combinations of non-Higgsable clusters, have been
observed both in 6D supergravity theories and 6D superconformal field
theories constructed from F-theory \cite{clusters, Heckman-Morrison-Vafa}.  In the
context of the toric bases we consider here, these can be seen as
arising simply from the sequences of primitive rays in a toric base
associated with a face of the bounding polytope at different distances
from the origin.  We encounter the $E_8$ sequence connecting $-12$ curves
in the example in \S\ref{sec:example-p2}, and  the simple SO(8)
sequence connecting $-4$ curves in the example in
\S\ref{sec:other-fiber}.  We briefly discuss here all the
constructions of this type to illustrate how they arise in a unifying
context in this framework.

\begin{figure}
\begin{center}
\includegraphics[width=8cm]{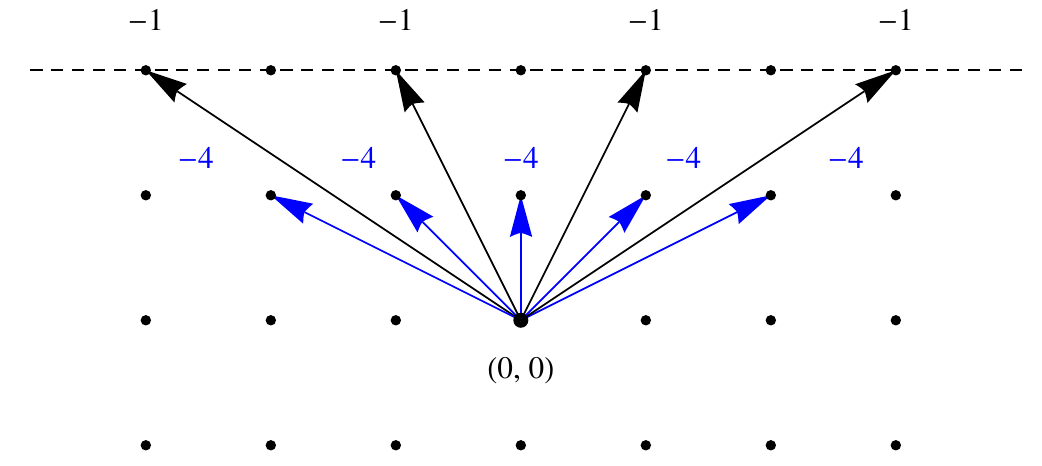}
\end{center}
\caption[x]{\footnotesize  Primitive rays associated with a face of
  the toric base polytope at distance 2 from the origin give a
  sequence of toric curves of self-intersection $-1, -4, -1, -4, \ldots$}
\label{f:distance-2}
\end{figure}

The simplest case is that of the self-intersection sequence $[[-1, -4,
-1, -4, -1, \ldots]]$.  This can be seen as arising from the set of
primitive rays associated with a face of the toric base polytope at
distance 2 from the origin (Figure~\ref{f:distance-2}).  The primitive
rays in this case are of the forms $(n, 1) \forall n$ and $(2k + 1, 2)
\forall k$.  For example, starting from $(0, 1)$, the sequence of rays
in a toric diagram associated with a polytope having a face along the
line $(x, 2)$ is
\begin{equation}
(0, 1), (1, 2), (1, 1), (3, 2), (2, 1), \ldots
\label{eq:rays-2}
\end{equation}
Since from toric geometry we know that a toric curve $v_i$ has
self-intersection $-n$ when $n v_i = v_{i -1} + v_{i +1}$, we can read
off the self-intersection sequence [[-4, -1, -4, -1, -4, \ldots]] from
the ray sequence (\ref{eq:rays-2}).  This corresponds to a sequence of
non-Higgsable gauge groups $SO(8), \cdot, SO(8), \cdot, SO(8) \ldots$
in the F-theory picture.

Performing a similar analysis at other distances we see that the
following sequences arise:
\begin{eqnarray}
d = 2 & \rightarrow & [[-4, -1, -4, \ldots]] \; (SO(8))\\
d = 3 & \rightarrow & [[-6, -1, -3, -1, -6, \ldots]]\; (E_6 \times SU(3))\\
d = 4 & \rightarrow & [[-8, -1, -2, -3, -2, -1, -8, \ldots]] \;(E_7
\times
(SU(2) \times SO(7) \times SU(2)))\\
d = 6 & \rightarrow & [[-12, -1, -2, -2, -3, -1, -5, -1, -3, -2, -2,
    -1, -12]] \; \nonumber\\ & &
\hspace*{0.3in}(E_8 \times F_4 \times (G_2 \times SU(2))^2)
\end{eqnarray}
where in each case the sequence repeats and we have indicated the
non-Higgsable gauge group for a single cycle of the sequence.  These
are precisely the maximal connected sequences identified in
\cite{clusters} and associated with e.g. ``$E_8$ matter'' in
\cite{Heckman-Morrison-Vafa}.

\end{document}